\documentclass[11pt]{article}
\usepackage{epsfig,cite}
\usepackage{graphicx}
\usepackage{multicol}
\usepackage{color}
\usepackage{amssymb}

\def\bea{\begin{eqnarray}}
\def\eea{\end{eqnarray}}

\newcommand\prd[3]   
		{{Phys.\ Rev.\ }{\bf D #1} (#2) #3}
\newcommand\prl[3]   
		{{Phys.\ Rev.\ Lett.\ }{\bf #1} (#2) #3}
\newcommand\plb[3]   
		{{Phys.\ Lett.\ }{\bf B #1} (#2) #3}
\newcommand\npb[3]    
		{{Nucl.\ Phys.\ }{\bf B #1} (#2) #3}
\newcommand\app[3]   
		{{Astropart.\ Phys.\ }{\bf #1} (#2) #3}
\newcommand\jhep[3]  
		{{J. High Energy Phys.\ }{\bf #1} (#2) #3}
\newcommand\epjc[3]  
		{{Eur.\ Phys.\ J. }{\bf C #1} (#2) #3}
\newcommand\npps[3]  
		{{Nucl.\ Phys.\ }{\bf #1} {\it(Proc.\ Suppl.)} (#2) #3}
\newcommand\jcap[3]  
		{{JCAP\ }{\bf #1} (#2) #3}

\newcommand\apj[3]    
		{{\it Astrophys.\ J. }{\bf #1} (#2) #3}

\newcommand{\hepex}[1]{{\tt hep-ex/#1}}

\newcommand{\gsim}{~{}_{\textstyle\sim}^{\textstyle >}~}
\newcommand{\lsim}{~{}_{\textstyle\sim}^{\textstyle <}~}

\newcommand{\CPV}{CP\!\!\!\!\!\!\!\!\raisebox{0pt}{\small$\diagup$}}

\begin{document}
\begin{titlepage}
\pagestyle{empty}
\baselineskip=21pt
\vspace*{2cm}
\begin{center}
{\huge\sf
Baryogenesis, Electric Dipole Moments and\\[0.3cm] Dark Matter in the MSSM}\\[0.4cm] 
\end{center}
\begin{center}
\vskip 0.6in
{\Large\sf  V.~Cirigliano, S.~Profumo and M.~J.~Ramsey-Musolf}\\
\vskip 0.2in
{\it {California Institute of Technology, Pasadena, CA 91125, USA}}\\
{E-mail: {\tt vincenzo@caltech.edu, profumo@caltech.edu, mjrm@caltech.edu}}\\
\vskip 0.4in
{\bf Abstract}
\end{center}
\baselineskip=18pt \noindent

\noindent We study the implications for electroweak baryogenesis (EWB) within the minimal supersymmetric Standard Model (MSSM) of present and future searches for the permanent electric dipole moment (EDM) of the electron, for neutralino dark matter, and for supersymmetric particles at high energy colliders. We show that there exist regions of the MSSM parameter space that are consistent with both present two-loop EDM limits and the relic density and that allow for successful EWB through resonant chargino and neutralino processes at the electroweak phase transition. We also show that under certain conditions the lightest neutralino may be simultaneously responsible for both the baryon asymmetry and the dark matter. We give present constraints on chargino/neutralino-induced EWB implied by the flux of energetic neutrinos from the Sun, the prospective constraints from future neutrino telescopes and ton-sized direct detection experiments, and the possible signatures at the Large Hadron Collider and International Linear Collider.

\vfill
\end{titlepage}
\baselineskip=18pt

\noindent\rule\textwidth{.1pt}	
\tableofcontents
\vspace*{0.5cm}
\noindent\rule\textwidth{.1pt}	

\vspace*{0.5cm}

\section{Introduction}

Despite the considerable successes of the Standard Model (SM) of particle physics in describing a wide array of experimental observations, various pieces of evidence --  ranging from neutrino masses and mixing to the abundance of baryonic and non-baryonic dark matter (DM)  in the universe -- point to physics beyond the SM. One of the most  widely explored scenarios for new physics is the supersymmetric extension of the Standard Model. While the  minimal formulation -- known as the MSSM -- does not address the question of neutrino mass, it  may resolve a number of other SM puzzles   by providing for  gauge coupling unification, the stability of the electroweak scale against large radiative corrections, the existence of a natural DM candidate (the lightest neutralino), and a mechanism for producing the observed baryon asymmetry of the universe (BAU) during the electroweak phase transition.  These potential remedies for SM shortcomings outweigh the costs involved in introducing supersymmetry, including the spectrum of as-yet unobserved superpartners of SM particles and the large number of {\em a priori} unknown parameters present even within the minimal model. Up-coming measurements at the Large Hadron Collider may discover the superpartners and -- together with precision electroweak measurements at both low-energies and the International Linear Collider -- may provide detailed information about the corresponding masses, mixing angles, and other parameters in the theory.

While many aspects of the MSSM have been exhaustively scrutinized, there has been recent, renewed interest in the CP-violating sector of the model. In particular, new searches for the permanent electric dipole moments (EDMs) of the electron, neutron, neutral atoms, and possibly the muon as well as deuteron ion, aim to improve the sensitivity beyond present experimental limits by up to four (or more) orders of magnitude (for recent reviews, see Refs.~\cite{Erler:2004cx,Pospelov:2005pr}). It is widely believed that if new CP-violating interactions at the electroweak scale are responsible for generating the BAU, then these new EDM searches would yield non-zero results. Present EDM limits already constrain the CP-violating phases in the MSSM to be quite small for slepton and squark masses $\lsim 1$ TeV, thereby tightly constraining (but not ruling out) the viability of MSSM electroweak baryogenesis (EWB). If nature is at least minimally supersymmetric, the future EDM measurements should either discover CP-violation at a level needed for successful EWB or rule it out altogether.

At the same time, a new generation of DM searches will have similarly significant implications for the viability of supersymmetric dark matter. Existing underground cryogenic detectors have already started to probe the spin-independent neutralino-proton scattering cross section for an increasingly large portion of the supersymmetric parameter space \cite{Munoz:2003gx}; the upgrade of existing (``Stage 2'') facilities (CDMS2\cite{cdms2}, Edelweiss2\cite{edelweiss}, CRESST2\cite{cresst}, ZEPLIN2\cite{zeplin}), and the forthcoming planned ``Stage 3'' ton-size detectors (XENON\cite{xenon}, GERDA\cite{gerda}, ZEPLIN4\cite{zeplin4} and WARP\cite{warp}) will eventually be sensitive to a wide array of supersymmetric dark matter models starting near the end of the decade. Further, neutralino annihilations in the core of the Sun, producing high-energy neutrino fluxes, will potentially give rise to signals large enough to overcome the atmospheric neutrino background in neutrino telescopes such as Antares\cite{antares} or IceCube\cite{icecube}, both currently in the deployment stage.

Theoretically, recent work suggest that -- apart from the CP-violating phases -- most MSSM parameters relevant for EWB are related to those that govern the properties of neutralinos and charginos. Thus, DM phenomenology can have important implications for the viability of MSSM EWB. In particular, the need for a strong first order phase transition to prevent washout of baryon number -- coupled with the restrictions from precision electroweak measurements -- indicate that the interactions of higgsinos and gauginos with the spacetime varying Higgs vacuum expectations values (vevs), rather than those involving scalar quarks and leptons, drive MSSM baryogenesis. Moreover, the effect of these interactions may become resonantly enhanced when one or other of the soft gaugino mass parameters, $M_{1,2}$ become comparable to the supersymmetric mass parameter $\mu$\cite{resonantewb,Lee:2004we}.  Under these conditions, MSSM EWB can be effective with the relatively small CP-violating phases that are required by one-loop EDM constraints for light sleptons. At the same time, the near degeneracy of these mass parameters leads to  a large degree of higgsino-gaugino mixing, a configuration yielding typically sizable DM direct and indirect detection rates. As the sensitivity of both DM and EDM searches improves, the results from both classes of experiments will substantially strengthen the constraints on MSSM EWB.

In this paper, we analyze the present and prospective implications of EDM and DM searches for baryogenesis in the MSSM. Our work is similar in spirit to the recent studies of Refs.~\cite{Balazs:2004bu,Balazs:2004ae}, but we make several observations beyond those contained in these analyses. In particular:

\begin{itemize}

\item[i)] From present experimental limits on the electric dipole moment of the electron, $d_e$, and one-loop calculations, we find that resonant, MSSM EWB is viable for sufficiently heavy sleptons. Moreover, there exists a range of slepton masses, $m_{\tilde\ell}$, for which one-loop contributions dominate $d_e$ but do not rule out resonant EWB. We determine the scale of $m_{\tilde\ell}$ for which two-loop contributions become comparable to one-loop effects. We find that two-loop effects already give important constraints on resonant MSSM EWB, especially with large CP violating phases and a light Higgs sector \cite{Pilaftsis:2002fe}; moreover, the anticipated level of the two-loop contributions within this scenario is predicted to fall well within the planned future experimental sensitivity on $d_e$ \cite{edmfuture}.

\item[ii)] When $\mu$ is comparable to the $U(1)_Y$ gaugino mass parameter $M_1$, resonant EWB in the MSSM can occur solely via interactions in the neutralino sector, whereas the authors of Refs.~\cite{Balazs:2004bu,Balazs:2004ae} concentrated on chargino-driven EWB. For this neutralino-driven EWB situation, the requirements of neutralino relic abundance and constraints from DM searches place additional constraints on resonant MSSM EWB. 

\item[iii)] Consistency between the DM relic abundance -- inferred from a wide array of cosmological and astrophysical data \cite{Spergel:2003cb} --  and neutralino-driven EWB implies a need for non-thermal DM production or cosmological enhancement mechanisms, whereas chargino-driven EWB can be consistent with thermal relic abundance considerations under some scenarios for SUSY-breaking mediation.

\item[iv)] The absence of energetic solar neutrinos in data collected by Su\-per\-Ka\-mio\-kan\-de places additional constraints on neutralino-driven EWB. Forthcoming ${\rm km}^3$-sized neutrino telescope and ton-sized direct DM detection experiments will probe most of the parameter space associated with both neutralino- and chargino-driven MSSM EWB. While the relic abundance requirements were studied for chargino-driven EWB in Refs.~\cite{Balazs:2004bu,Balazs:2004ae}, we provide a comprehensive study of both relic abundance and DM detection implications for MSSM EWB. Moreover, we consider the phenomenology of both minimal supergravity (mSUGRA) as well as anomaly-mediated SUSY breaking (AMSB) scenarios, whereas the authors of Ref.~\cite{Balazs:2004bu,Balazs:2004ae} concentrated on the mSUGRA case. 

\item[v)] We give a comprehensive review of the prospects for the discovery of MSSM EWB at colliders, including the Tevatron, the Large Hadronic Collider (LHC) and a future International Linear Collider (ILC), outlining the regions, compatible with the generation of the BAU and with dark matter, which will fall within the anticipated sensitivity of the various accelerator facilities.

\end{itemize}

In the reminder of the paper we explore the resonant MSSM EWB scenario (Sec.~\ref{sec:mssm}), outlining the viable portions of the parameter space (Sec.~\ref{sec:mssmparamspace}), and studying the wide array of resulting phenomenological consequences. In particular, we consider the implications of EDM searches  (Sec.~\ref{sec:edmconst}), the DM relic abundance (Sec.~\ref{sec:omega}), current constraints and future prospects for direct and indirect DM detection (Sec.~\ref{sec:dmsearches}), and the role of collider searches for supersymmetric particles (Sec.~\ref{sec:colliders}).

\section{Baryogenesis and Electric Dipole Moments  in the MSSM}\label{sec:mssm}

In principle, the SM itself contains all the ingredients  necessary for the generation of a BAU, given by the well known  Sakharov requirements \cite{sak}. It was demonstrated long ago, however,  that the resulting SM BAU is several orders of magnitude smaller than the observed value \cite{noewb}. While the gauge sector of the SM provides the necessary baryon number violation via SU(2)$_L$ sphaleron processes, it does not allow for a
strong first order EW phase transition (EWPT) as required to prevent washout of sphaleron-induced baryon number production\cite{Bochkarev:1987wf}, nor does it give rise to sufficiently large CP-violating effects needed to generate an imbalance of chiral charges that drive the sphaleron transitions. In the case of the EW phase transition,  one must satisfy the condition $v(T_c)/T_c\gtrsim1$, where $v$ is the Higgs vev at the critical transition temperature, $T_C$.  In the SM, the LEP II lower bound on the Higgs boson mass, $m_h\gsim 114$ GeV \cite{leph} prevents this condition from being satisfied. Similarly,  CP-violating processes in SM baryogenesis are highly suppressed by both the Jarlskog invariant and powers of the quark Yukawa couplings. 

The MSSM addresses both of these shortcomings.  Specifically, 
loop-induced  contributions to the MSSM  Higgs  potential from the third generation scalar quarks that have large Yukawa couplings and carry six degrees of freedom lead to a strong first order EWPT for values of the lightest Higgs mass greater than the LEP II lower bound as long as one of the top squarks is lighter than the top quark \cite{ewpt_perturbative,Carena:1997ki} (non-perturbative analyses also confirm this picture \cite{ewpt_nonperturbative}). As a corollary, one obtains an indirect upper bound on the mass of the lightest supersymmetric particle (LSP) --  assumed in this work  to be the lightest neutralino, hereafter indicated as $\chi$:  $m_\chi\lesssim m_t$. On the other hand, the MSSM EWPT becomes too weakly first order for $m_h\gsim 120$ GeV, leaving a rather small window for viable EWB (for recent studies of these considerations, see {\em e.g.}, Refs.~\cite{resonantewb,cline,mssmewb,Carena:1997ki,Carena:2002ss}.) It is important to note, however, that extensions of the MSSM with an enlarged Higgs sector or additional U(1) symmetries can relax these requirements on the Higgs and top squark masses, thereby opening a larger window for a strong first order EWPT and allowing for a more massive $\chi$ \cite{nonminimal}. 

The effects of CP-violation enter as source terms in the quantum transport equations that govern the production of chiral charge at the phase boundary. Chiral charge production depends on a detailed competition between CP-odd particle-antiparticle decay and scattering asymmetries, CP-conserving charge relaxation processes as favored by free energy minimization, and transfer of number density from one species to another through various reaction mechanisms. To illustrate, we consider the transport equation for combined Higgs/higgsino number density \cite{Cirigliano:2006wh}:
\begin{equation}
\label{eq:Heq}
\partial^\mu H_\mu = - \Gamma_H\frac{H}{k_H}
-\Gamma_Y\biggl(\frac{Q}{k_Q} - \frac{T}{k_T} + \frac{H}{k_H}\biggr) -
{\tilde\Gamma}_Y\biggl(\frac{B}{k_B} - \frac{Q}{k_Q} +
\frac{H}{k_H}\biggr)+ \bar\Gamma_Y\frac{h}{k_h} + S_{\widetilde
H}^{\CPV} \ \ \ , 
\end{equation} 
Here, $H$ and $h$ are number densities associated various combinations
of the up- and down-type Higgs supermultiplets in the MSSM (defined
below); $H_\mu$ is the corresponding vector current density; $Q$ and
($B$,$T$) are the number densities of particles in the third
generation left- and right-handed quark supermultiplets, respectively;
the $k_{H,h,Q,T,B}$ are statistical weights; $S_{\widetilde H}^{\CPV}$
is a CP-violating source; and $\Gamma_H$, $\Gamma_Y$,
${\tilde\Gamma}_Y$, and $\bar\Gamma_Y$ are transport coefficients associated with relaxation of Higgs supermultiplet densities and their transfer to the baryon sector. 
Analogous expressions hold for the quark supermultiplet densities, $Q$, $T$, {\em etc}. In the SM, the CP-violating source term 
$S_{\widetilde H}^{\CPV}$ vanishes, whereas the analogous terms $S_{Q,T,\ldots}^{\CPV}$ for quark densities are suppressed by the factors indicated above. In the MSSM, however, $S_{\widetilde H}^{\CPV}$ is generally non-zero and is proportional to the relative phase between $\mu$ and the gaugino mass parameters, $M_{1,2}$. Similarly, the $S_{Q,T,\ldots}^{\CPV}$ depend on both ${\rm arg}(\mu M_{1,2})$ and the relative phase between $\mu$ and the triscalar couplings, $A_f$. In general, these terms contain no suppression factors other than the CP-violating phases that are constrained by EDM measurements, as discussed below. 

Solving the coupled set of transport equations and the corresponding baryon number diffusion equation leads to a simple expression for the baryon-to-entropy ratio, $Y_B$ \cite{Lee:2004we}:
\begin{equation}
\label{eq:yb}
Y_B\equiv\frac{n_B}{s} = F_1\, \sin\phi_\mu + F_2\, \sin(\phi_\mu+\phi_A)\ \ \ ,
\end{equation}
where we have taken the $M_{1,2}$ to be real and have assumed a common phase $\phi_A$ for the soft breaking triscalar couplings. The coefficients $F_i$ depend on the other mass parameters in the MSSM as well as on characteristics of the expanding bubbles of broken EW symmetry.  
The first term in Eq.~(\ref{eq:yb}) arises from CP-violating processes in the higgsino-gaugino sector, while the second term is generated by squark CP-violation. In each case, the the $F_i$ depend roughly on the strength of the CP-violating source in a given sector divided by the square root of an average relaxation coefficient as well as on the rates for particle diffusion ahead of the bubble wall ($\Gamma_{\rm diff}$),  sphaleron transitions ($\Gamma_{\rm sph}$), and particle density transfer ($\Gamma_Y$). In the $\Gamma_Y\to\infty$ limit, one has roughly
\begin{equation}
\label{eq:scaling}
F_i\sim \frac{S_i^{\CPV}}{\sqrt{\bar\Gamma}}\, \frac{\Gamma_{\rm sph}}{\Gamma_{\rm diff}} \ \ \ .
\end{equation}
Corrections to the $\Gamma_Y\to\infty$ limit have recently been investigated in Ref.~\cite{Cirigliano:2006wh}.

For $\mu$ not too different from either $M_1$ or $M_2$, or for $m_{\tilde Q}$ close to $m_{\tilde t_R}$, both the CP-violating sources \cite{cline,resonantewb} and the relaxation coefficients \cite{Lee:2004we} can become resonantly-enhanced.  The net effect is an enhancement of one or the other of the $F_i$ in Eq.~(\ref{eq:yb}), thereby allowing effective EWB with smaller CP-violating phases. Although there is overall agreement in the literature about the occurrence of the aforementioned resonant enhancement, its precise numerical magnitude is a matter of continued investigation. In particular, different treatments of the source terms that enter the quantum transport equations lead to some numerical differences. In Ref.~\cite{Lee:2004we} -- upon which we rely in the present study -- the sources were computed using the closed time path (CTP) methods, a basis of weak eigenstates for the superpartners, and a ``mass insertion'' approximation that is valid in the domain of relatively small and gently varying Higgs vevs. Both the CP-violating sources as well as relaxation terms and Higgs-baryon transport coefficients were computed in a self-consistent manner, including various leading-order effects not previously included in other studies. The numerical results are similar to those obtained by Carena et al., \cite{resonantewb,Carena:2002ss}, who computed the CP-violating sources using the similar CTP methods, included higher order contributions in derivatives of the Higgs vevs, but did not compute the other transport coefficients within the same framework. In contrast, the authors of Ref.~\cite{Konstandin:2005cd} used a basis of local mass eigenstates for the CTP computation of the CP-violating sources but a phenomenological model for the relaxation rate. These authors find a somewhat smaller baryon asymmetry compared to Refs.~\cite{Lee:2004we,resonantewb,Carena:2002ss}. In view of other ${\mathcal O}(1)$ uncertainties in all present baryogenesis computations, we do not consider these numerical differences to be significant.

The resonant effect is phenomenologically disfavored in the squark sector by competing requirements of a strong first order phase transition and precision electroweak measurements. The former requires a light stop quark while the latter require the LH stop to be much heavier the the top quark. Thus, consistency between MSSM EWB and the small CP-violating phases implied by current EDM limits and one-loop MSSM contributions point toward the resonant scenario in the higgsino/gaugino sector. 

Here, we analyze the dependence of these resonant processes on the mass parameters in this sector of the theory. In doing so, we vary $|\mu|$, $M_{1,2}$, and their relative phases while choosing other parameters -- including heavy Higgses mass scale, the stop mass parameters,  and $\tan\beta$, the ratio of the MSSM Higgs vacuum expectation values -- in order to satisfy  the requirements of a strong first order EWPT and  consistency with precision electroweak data \cite{resonantewb,cline,mssmewb,Carena:1997ki,Carena:2002ss}. For instance, the heavier stop must be at or above the TeV scale, the stop mixing parameter must be sufficiently small ($|A_t-\mu/\tan\beta|\lesssim0.5 M_{Q_3}$), and $\tan\beta$ must be larger than 5 \cite{Carena:1997ki}. Further, in order to avoid unacceptably large contributions to the $\rho$ parameter, the lightest stop must be mostly right-handed. The BAU in the MSSM also depends critically on $m_A$, the mass scale of the CP-odd Higgs. In particular, increasing $m_A$ yields a significant suppression of the relative variation of the two Higgs fields along the bubble walls, $\Delta\beta$ -- a parameter upon which the BAU depends linearly \cite{Moreno:1998bq}. As far as the CP violating (CPV) phases are concerned, we consider here only the relative gaugino-higgsino phase $\phi_\mu\equiv{\rm Arg}(\mu M_2)={\rm Arg}(\mu M_1)$  and disregard other possible non-trivial CPV phases\footnote{The equality of gaugino phases follows from our assumptions on the gaugino spectrum as discussed below.}.

\begin{table}[!t]
\begin{center}
\begin{tabular}{|c|c|c|c|c|c|}\hline
$\tan\beta$&$m_{\widetilde U_3}$&$m_{\widetilde Q_3}$&$A_t$&$m_{\widetilde F}$&$m_A$\\
\hline
10&90 GeV&10 TeV&$0.4\ m_{\widetilde Q_3}$&10 TeV&150 GeV; 1 TeV\\
\hline
\end{tabular}
\end{center}
\caption{\it\small The MSSM setup under investigation here; $m_{\widetilde F}$ refers to the generic soft breaking scalar masses, other than those otherwise specified; the other trilinear scalar couplings were set to 0.}\label{tab:mssm}
\end{table}
The above remarks lead us to consider the MSSM reference scenario summarized in Tab.~\ref{tab:mssm}. In this framework, the resulting lightest stop mass is close to the top mass, and the Higgs mass is around 118 GeV. While the Higgs mass can be lowered down to the LEP2 limit without affecting significantly any of the quantities of interest here (for instance tuning $A_t$), lower lightest stop masses would further shrink the available parameter space, forcing the LSP mass to be even lighter (we recall that $m_\chi<m_{\widetilde t_1}$); we therefore regard this reference scenario as a conservative one, and the parameter space we consider as the virtually maximal one expected in the context of EWB in the MSSM.

As far as the pattern of soft, SUSY-breaking gaugino masses is concerned, 
we concentrate on two well-motivated choices:

\begin{itemize}

\item[(1)] In the context of gravity-mediated supersymmetry breaking, assuming a trivial gauge kinetic function, we consider the standard GUT-scale unification of the gaugino masses to a common value $M_{1/2}$; this choice leads -- after renormalization group evolution from the GUT scale down to the electroweak scale -- to the approximate gaugino mass pattern \cite{Baer:2000gf}
\begin{equation}
M_1:M_2:M_3\approx 1:2:6\quad{\rm GUT-scale\ gaugino\ mass\ unification}.
\end{equation}

\item[(2)] Anomaly-mediated supersymmetry breaking contributions may dominate the soft-supersymmetry breaking gaugino masses $M_i$ \cite{Randall:1999ee,Moroi:1999zb}, yielding the relation
\begin{equation}
M_i= \frac{\beta_{g_i}}{g_i}\ m_{3/2},
\end{equation}
where the symbols $g_i$ stand for the gauge couplings, $\beta_{g_i}$ for the relative one-loop $\beta$-functions, and $m_{3/2}$ for the gravitino mass. In this case, one has the following mass pattern \cite{Moroi:1999zb,Baer:2000gf}:
\begin{equation}
M_1:M_2:M_3\approx 3:1:8\quad{\rm anomaly\ mediation}.
\end{equation}
In this latter case, the usual hierarchy $M_1\ll M_2$ is inverted, giving rise to a peculiar neutralino/chargino spectrum featuring -- in the limit of large $\mu$ --  a wino-like LSP with an almost degenerate lightest chargino, and, consequently, a very distinctive related phenomenology \cite{Moroi:1999zb}. 

\end{itemize}
Since the critical mass parameters in resonant, higgsino/gaugino-dominated EWB are the gaugino masses and the higgsino mass term $\mu$, we concentrate on $(M_{1},\mu)$-dependence of $Y_B$ for GUT-scale gaugino mass unified models and the $(M_{2},\mu)$-dependence for the anomaly mediated gaugino mass pattern case.

The foregoing parameters also govern supersymmetric contributions to the flavor-changing decays
$b\rightarrow s\gamma$ and $B_s\rightarrow\mu^+\mu^-$ as well as the muon anomalous magnetic moment. In the latter two cases, the large values of the relevant sfermion masses and the low $\tan\beta$ value we use here render the supersymmetric contributions completely negligible.
The implications of $b\rightarrow s\gamma$ are potentially more significant.
We have computed the constraints on the parameter space under consideration, assuming a maximal CP violating phase $\phi_\mu=\pi/2$. We find that the largest contributions typically come from the charged Higgs loop diagrams, and are particularly significant at low values of $m_A$. The $m_A=1000$ GeV case is safely within the range allowed by the theoretical uncertainties from the SM computation \cite{bsgth} and by the experimental error \cite{bsgexp} for the whole range of higgsino and gaugino masses under consideration here. Setting $m_A=150$ GeV, instead, typically gives rise to values very close to, or even in excess of, the 2-$\sigma$ upper limit on  BR($b\rightarrow s\gamma$), depending on how the experimental and theoretical errors are combined. Adding the theory and experimental errors in quadrature favors values of $m_A\gtrsim200$ GeV, while adding them linearly leaves the freedom of allowing values of $m_A$ as low as 150 GeV. With the purpose of highlighting extreme options for the EWB scenario, we will here consider the $m_A=150$ GeV case, bearing in mind that this would entail potentially large contributions to $b\rightarrow s\gamma$.

\subsection{EWB and the supersymmetric parameter space}\label{sec:mssmparamspace}

We begin our numerical analysis by identifying the regions of the MSSM parameter space compatible with the value of $Y_B$ determined from the cosmic microwave background \cite{Spergel:2003cb}:
\begin{equation}
\label{eq:ybwmap}
Y^{\rm WMAP}_B\ =\ (9.2\pm1.1)\times10^{-11}, \quad {\rm WMAP}.
\end{equation}
As discussed above, we concentrate on the planes defined by the supersymmetric higgsino mass term $\mu$ and one of the soft SUSY breaking gaugino masses $M_{1,2}$, the other one being fixed by assumptions on the mechanism of SUSY breaking. Since a more complicated dependence links  $Y_B$ and $m_A$ through the parameter $\Delta\beta$, we consider the limiting case $m_A=150$ GeV. Larger values of $m_A$ will suppress $Y_B$, and the resulting $Y_B$ contours can be read from what we show here through a proper $m_A$-dependent rescaling factor \cite{Moreno:1998bq}. As far as the bubble wall parameters are concerned, consistently with the supersymmetric parameters choice of Tab.~\ref{tab:mssm}, we adopt the central values $v_w=0.05$ \cite{wallvelocity} and $L_w=25/T$ \cite{Moreno:1998bq}, and the parameterization of $\Delta\beta$ as a function of $m_A$ provided in Ref.~\cite{Moreno:1998bq}. We have made no attempt to include a theoretical error bar  on the computation of $Y_B$, since various approximations that enter the calculation remain to be scrutinized\cite{Lee:2004we,Cirigliano:2006wh}. One should, therefore, treat the allowed regions as indicative of favored values of the MSSM parameters rather than as airtight limits.

\begin{figure}[!t]
\begin{center}
\hspace*{-1.cm}\epsfig{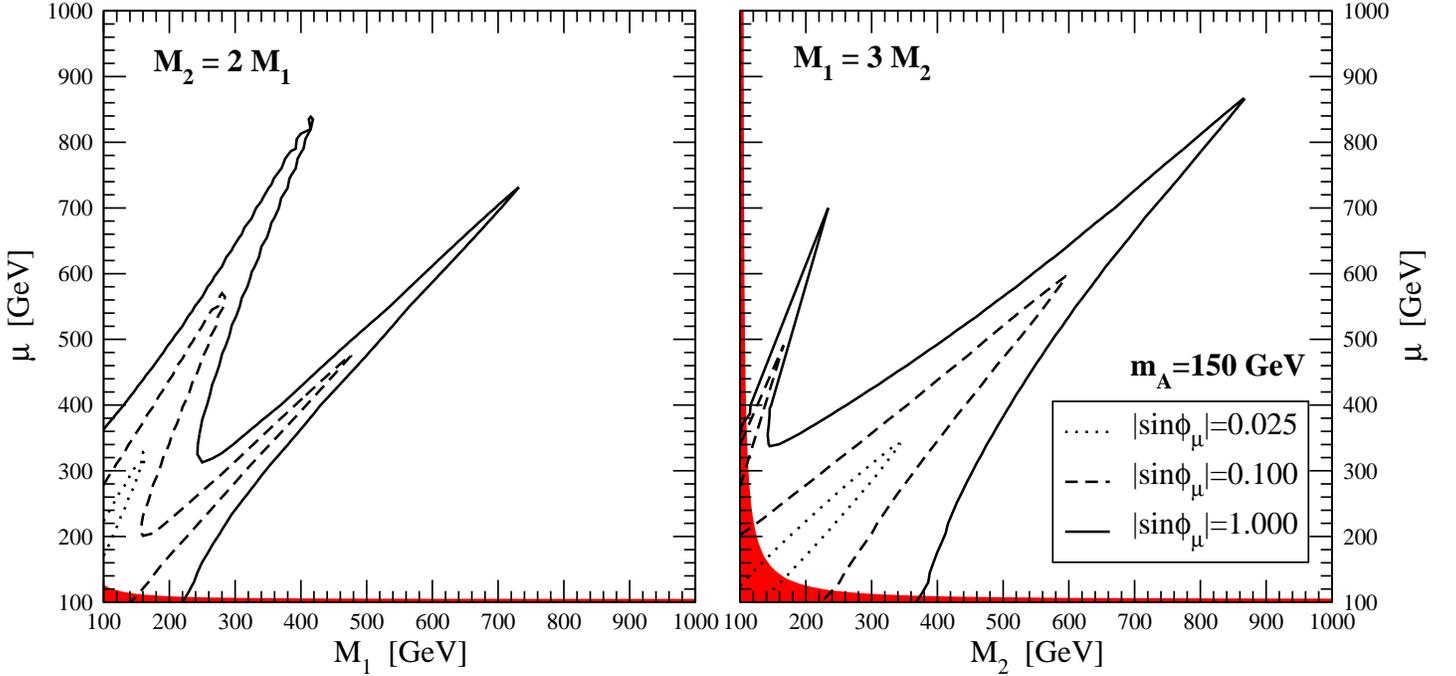}
\end{center}
\caption{\it\small 
Regions, on the $(M_{1,2},\mu)$ planes, producing the central value for the baryon asymmetry in the Universe deduced from the WMAP analysis \cite{Spergel:2003cb}, for various values of the CP-violating phase $\phi_\mu$, at $m_A=150$ GeV. In panel {\em (a)} we assume gaugino soft breaking masses unification at the GUT scale, leading, at the EW scale, to $M_2\simeq 2M_1$, while in panel {\em (b)} we assume an anomaly-mediated inspired gaugino mass pattern, leading to $M_1\simeq 3M_2$. The red regions correspond to chargino masses below the LEP2 limit ($m_{\widetilde\chi^\pm}<103.5\ {\rm GeV})$.}
\label{fig:yb}
\end{figure}

Fig.~\ref{fig:yb} shows the $Y_B^{\rm th}=Y^{\rm WMAP}_B$ curves (where $Y_B^{\rm th}$ indicates the theoretically computed BAU) on the ($M_i,\mu$) planes, for various values of $\sin\phi_\mu$ at $m_A=150$ GeV (the corresponding curves at $m_A=1$ TeV, a value we will use later in our analysis, follow a trivial rescaling by a suppression factor $\approx15$ \cite{Moreno:1998bq}). In panel {\em (a)} we assume gaugino mass unification at the GUT scale, while in {\em (b)} we assume the gaugino spectrum expected in the case of anomaly mediation of SUSY breaking. The red (dark grey) regions indicate the (conservative\footnote{For small mass differences between the lightest chargino and the LSP, the chargino mass bound is slightly weaker than the value we use here \cite{lep2}.}) LEP2 bound on the lightest chargino mass \cite{lep2}: points within the red-shaded portion of the plots feature $m_{\widetilde \chi^\pm_1}<103.5$ GeV. We extend here the range of explored parameters beyond the $m_\chi\lesssim m_t$ limit, assuming a heavier lightest stop, and contemplating the possibility that other, non-minimal mechanisms lead to the required sufficiently strong first order EW phase transition \cite{nonminimal} (but still assuming that the dominant source for the generation of the baryon asymmetry stems from the gaugino/higgsino sector). The curves in Fig.~\ref{fig:yb} lead to several observations:

\begin{itemize}

\item[(i)] As expected,  higgsino/gaugino-mediated EWB is consistent with $Y_B$
as long as one is close to a resonance, {\em i.e.} if one of the two gaugino masses is almost degenerate with $\mu$. This leads to the double-funnel structure in the plots: the heavier the neutralinos/charginos, the smaller the resulting $Y_B$ and the closer $\mu$ has to be to either $M_1$ or $M_2$. In all cases, we expect a non-negligible higgsino-gaugino mixing in the lightest neutralino, which is maximized along the $M_{1,2}\sim\mu$ line, where $Y_B$ is also maximal, particularly in the lower branches, where $\mu$ is degenerate with the lighter gaugino mass. 

\item[(ii)] The CP-violating higgsino term is effective enough within EWB in the MSSM as long as $\mu$ and at least one of the gaugino masses are smaller than roughly 1 TeV, conservatively taking into account potential effects of $\mathcal O$(1) in the evaluation of $Y_B$. Further, we stress that the values of $\sin\phi_\mu$ compatible with a large enough $Y_B$, even on top of the resonance, have to be larger than $\approx 10^{-2}$.

\item[(iii)] Not surprisingly, the $M_2\sim\mu$ resonance produces larger values of $Y_B$ than the $M_1\sim\mu$ case, since the bino-higgsino resonance cannot proceed through chargino exchange. Numerically, however, we find that the relative magnitudes of the two resonances are not substantially different:  ${Y_B (\mu \sim M_1)}/{Y_B (\mu \sim M_2)}\approx0.5$. This result suggests the possibility that neutralinos alone may drive MSSM EWB -- a possibility that was not considered in earlier studies and that has implications for supersymmetric DM (see below).

The relative magnitude of the $\mu\sim M_2$ and $\mu\sim M_1$ resonant peaks for $Y_B$ can be understood by referring to Eq.~(\ref{eq:scaling}) and noting that the strength of the CP-violating sources are governed by the gauge couplings and plasma damping rates $\Gamma_P^{{\tilde H},\ {\tilde W}}$ for the higgsinos and gauginos that resonantly scatter from the spacetime varying Higgs vevs\cite{Lee:2004we}:
\begin{eqnarray}
\nonumber
S_{{\tilde H}-{\tilde B}}^{\CPV} & \sim& g_1^2 \ \times \  \left(\Gamma_P^{\tilde H}+\Gamma_P^{\tilde B}\right) \\
\nonumber
S_{{\tilde H}-{\tilde W}}^{\CPV} &\sim & 3g_2^2 \ \times \  \left(\Gamma_P^{\tilde H}+\Gamma_P^{\tilde W}\right)\ \ \ 
\end{eqnarray}
so that
\begin{equation}
\frac{Y_B (\mu \sim M_1)}{Y_B (\mu \sim M_2)} \simeq \frac{g_1^2}{3
g_2^2} \times \left[ \frac{\Gamma_H (\mu \sim M_2)}{\Gamma_H (\mu \sim
M_1)} \right]^{1/2} \times \frac{ \Gamma^{\tilde{H}}_P +
\Gamma^{\tilde{W}}_P}{ \Gamma^{\tilde{H}}_P + \Gamma^{\tilde{B}}_P}  \  . 
\end{equation}
The first factor simply reflects the different number and nature of
the gaugino intermediate states in the two cases, and is approximately
equal to 0.1.  The second factor involves the ratio of the
Higgs-violating rates $\Gamma_H$ (see Eq.~\ref{eq:Heq}): the larger $\Gamma_H$, the smaller
the resulting baryon asymmetry. Numerically this is roughly two.
Finally, the third factor involves the ratio of the damping rates of
higgsinos and gauginos, arising from resonant energy denominators in
the expressions of Ref.~\cite{Lee:2004we}. Using the damping rates
calculated in Ref.~\cite{Elmfors:1998hh} we find this factor is again
approximately two, leading to $Y_B (\mu \sim M_1)/Y_B (\mu \sim M_2)
\sim 0.44$, which closely approximates what we find in the numerical 
evaluation.

\end{itemize}

Before discussing the impact of EDM searches on the scenarios in Fig.~\ref{fig:yb}, we note that extensions of the MSSM with enlarged Higgs sectors or additional U(1) symmetries\cite{nonminimal} will lead to modifications of both the CP-violating sources and relaxation terms entering the computation of $Y_B$. Thus, we emphasize that the constraints on the viability of supersymmetric EWB discussed here are specific to the MSSM.

\subsection{EDM constraints}\label{sec:edmconst}

A natural consequence of non-CKM CP violating phases in the MSSM, as required by EWB, is the generation of possibly large electric dipole moments (EDMs) through loops involving supersymmetric particles (for recent reviews, see Refs.~\cite{Erler:2004cx,Pospelov:2005pr}).
In general, the EDMs of different systems, such as charged leptons, the neutron, or neutral atoms, carry complementary dependences on the CP-violating phases in a given extension of the SM. Consequently, a comparison of results form different EDM searches provide a substantially more powerful probe of new CP-violation than the results from any single search alone. However, the implications of EDM searches for MSSM EWB are an exception to this general statement, since both resonant gaugino-higgsino EWB and the EDM of the electron, $d_e$, depend on a single CP-violating phase, $\phi_\mu$. Consequently, we focus here on constraints imposed by $d_e$ experiments. 

\begin{figure}[!t]
\begin{center}
\hspace*{-1.cm}\epsfig{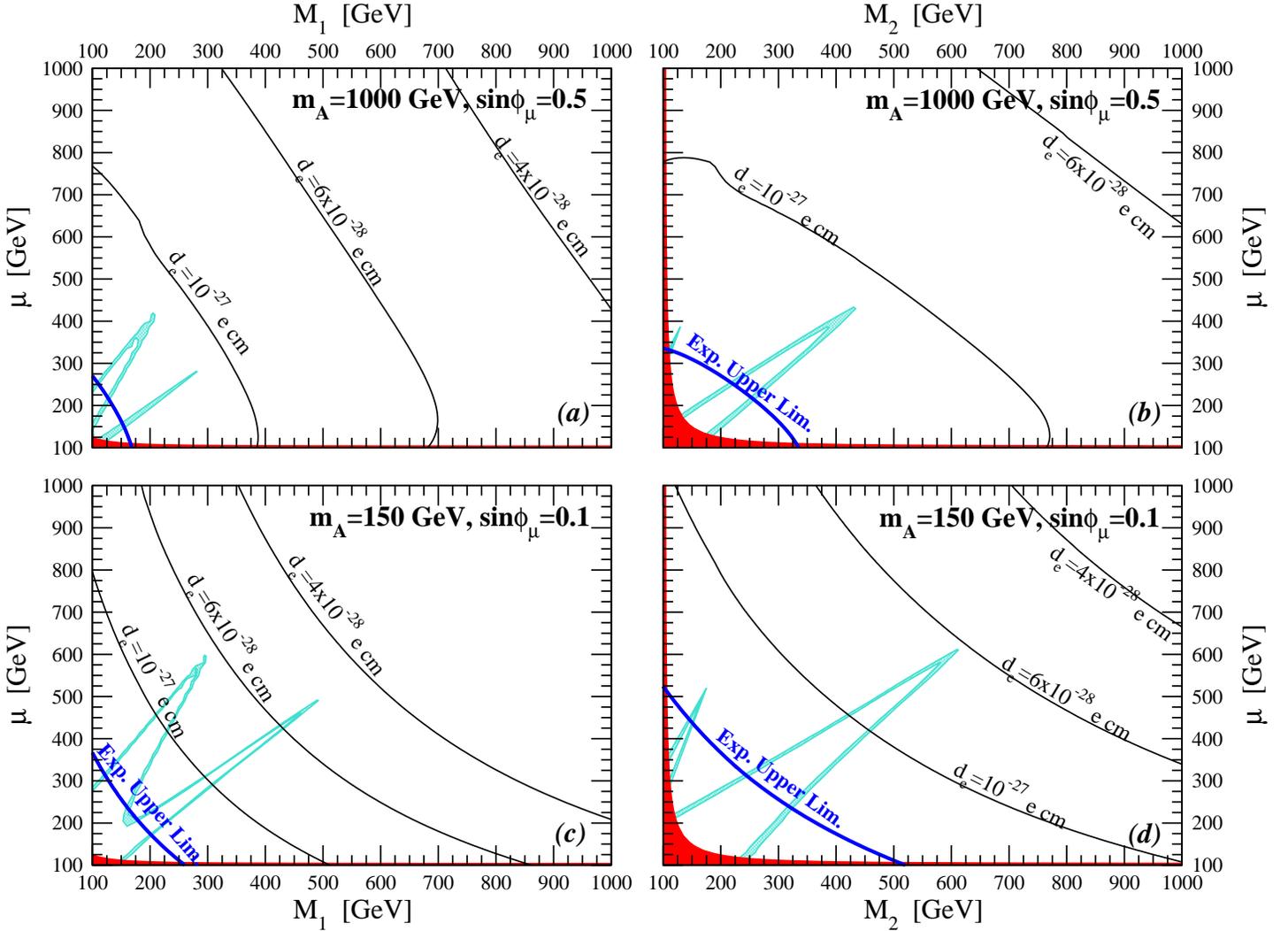}
\end{center}
\caption{\it\small 
Iso-level curves of the electron electric dipole moment, on the $(M_{1,2},\mu)$ planes, at various values of $m_A$ and of the CP violating phase $\phi_\mu$, in the limit of heavy sfermions. The thick blue lines represent the current experimental upper limit \cite{Regan:2002ta}. We also shade in light blue the 2-$\sigma$ regions corresponding to a WMAP BAU \cite{Spergel:2003cb}. In the two upper panels we take $m_A=1$ TeV and $\sin\phi_\mu=0.5$, while in the two lower panels $m_A=150$ GeV and $\sin\phi_\mu=0.1$. In panels {\em (a)} and {\em (c)} we assume gaugino soft breaking masses unification at the GUT scale, leading, at the EW scale, to $M_2\simeq 2M_1$, while in panels {\em (b)} and {\em (d)} we assume an anomaly-mediated inspired gaugino mass pattern, leading to $M_1\simeq 3M_2$. The red (dark) regions correspond to chargino masses below the LEP2 limit ($m_{\widetilde\chi^\pm}<103.5\ {\rm GeV})$.}
\label{fig:edm}
\end{figure}

For  values of the sfermion masses $\lsim$ 1 TeV, the dominant supersymmetric contributions to the EDMs of SM fermions originate from one-loop diagrams involving a sfermion and a supersymmetric fermion, such as charginos, neutralinos or gluinos. For larger sfermion masses, contributions from two-loop diagrams that contain supersymmetric fermions and Higgs and/or gauge bosons can give comparable contributions (see {\em e.g.} \cite{Pilaftsis:2002fe,Chang:2002ex}). In the case of the electron EDM, the largest two-loop contributions stem from diagrams involving a photon and a Higgs boson $\sigma=h,H,A$. We also include here (i) the subdominant contribution from $W^+W^-$ recently analyzed in \cite{Giudice:2005rz}, giving in the present context corrections on the order of the percent, and (ii) the contribution from $\widetilde t$-$\gamma$-$\sigma$ loops \cite{Chang:1998uc} (we recall that the lightest stop mass is here assumed to be at the top mass scale), which, while subdominant at small $\mu$, contributes to the level of a few percent at $\mu\sim1$ TeV.

Interestingly, we find that for maximal CPV phases, $\sin\phi_\mu\simeq1$, all the supersymmetric parameter space compatible with EWB is ruled out by the current experimental limit on $d_e\lesssim1.9\times10^{-27}$ e-cm (95\% C.L.) \cite{Regan:2002ta}, even in the limit of super-heavy sleptons, due to pure two-loop supersymmetric contributions to the electron EDM. In Fig.~\ref{fig:edm}, we show the two-loop EDM constraints in the same $(M_{1,2},\mu)$ 
planes studied in Fig.~\ref{fig:yb}, at various values of $m_A$ and of the CP violating phase $\phi_\mu$, to illustrate that EWB and the electron EDM are in general compatible. The thin black contours correspond to different values of $d_e$, while the dark blue contours trace the present experimental upper bound. The heavy Higgs contributions are greatly enhanced at lower values of $m_A\simeq m_H$. While both the BAU and $d_e$ are approximately proportional to $\sin\phi_\mu$, the scaling with $m_A$ is highly non-trivial. In general, the electron EDM constraint enforces large values for the heavy Higgs boson mass scale in the presence of large values of the CP-violating phase $\phi_\mu$, while at smaller CP violating phases EWB is still viable provided $m_A$ is sufficiently light. We find that the expected range of values for $d_e$ from two-loop EDM on the supersymmetric parameter space compatible with EWB is, in general, larger than $10^{-28}$ e-cm. Since the expected future sensitivity of $d_e$ searches might be as good as $\approx 10^{-29}\div 10^{-30}$ e-cm \cite{edmfuture}, one could expect observation of a non-zero $d_e$ if MSSM EWB is indeed the mechanism for the generation of the BAU. 

\begin{figure}[!t]
\begin{center}
\hspace*{-1.cm}\epsfig{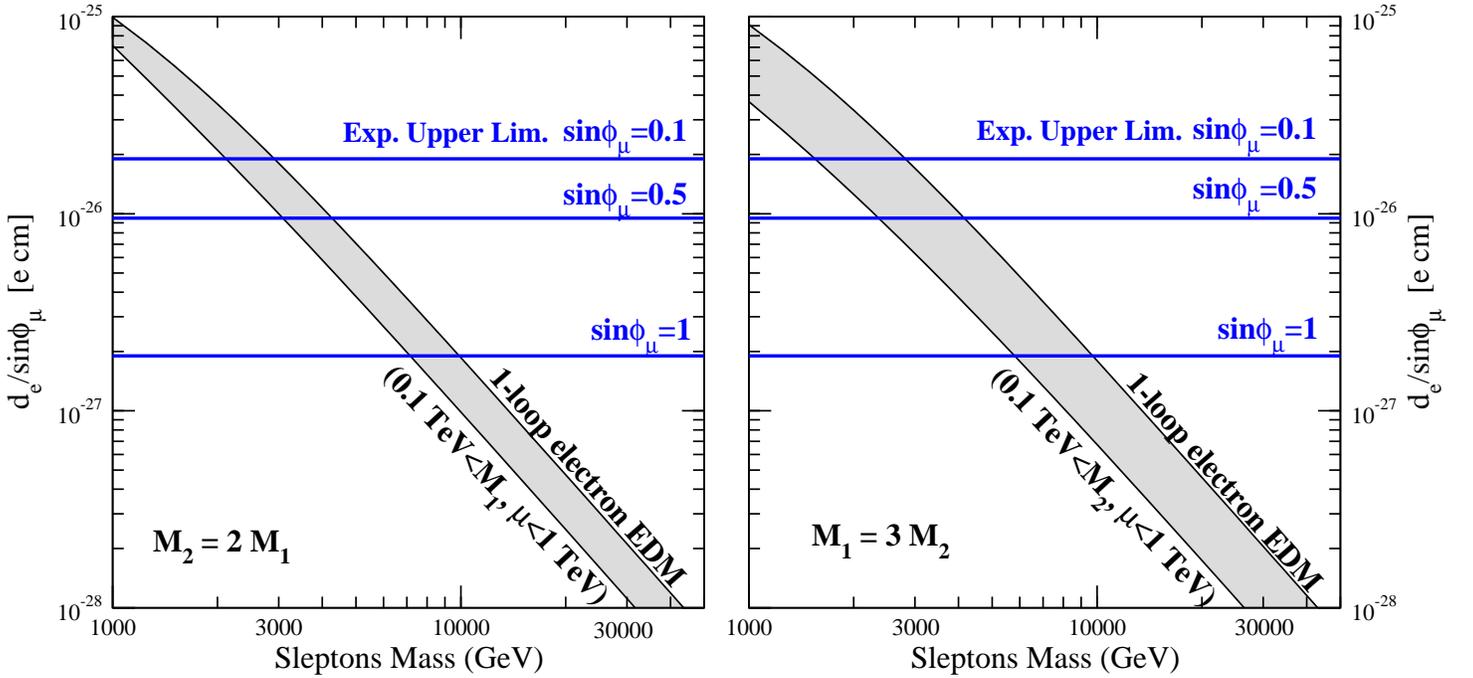}
\end{center}
\caption{\it\small The range of the one-loop electron EDM, as a function of the common slepton mass, obtained varying the gaugino soft breaking SUSY masses and the $\mu$ parameter on the planes shown in Fig.~\ref{fig:edm}, {\em i.e.} between 0.1 and 1 TeV. We plot the quantity $d_e/\sin\phi_\mu$, and show the experimental upper limit for various values of $\sin\phi_\mu$, including those shown in Fig~\ref{fig:edm}..}
\label{fig:1ledm}
\end{figure}
It is interesting to determine the conditions under which the one- and two-loop $d_e$ contributions have similar magnitudes and the values of the MSSM for which they can be compatible with present experimental limits. To that end, we show in Fig.~\ref{fig:1ledm} the one-loop electron EDM as a function of the slepton mass \cite{ibranath}; we assume all the relevant slepton sector (selectrons and electron sneutrino) to be degenerate, and for each given slepton mass we indicate with the grey band the range of variation of $d_e/\sin\phi_\mu$ over the planes in Fig.~\ref{fig:yb}. We also indicate with horizontal lines the current experimental upper limit for various values of $\sin\phi_\mu$, including those shown in Fig.~\ref{fig:edm}. We observe that consistency with the current experimental limit forces the slepton masses to lie above 5-10 TeV, with maximal CP violating phases. TeV sfermions imply $\sin\phi_\mu\lesssim0.02$, a value which Fig.~\ref{fig:yb} indicates to be barely consistent with EWB. This means that electron EDM constraints and successful MSSM EWB imply slepton masses larger than a TeV. 

We also observe that, independently of $\phi_\mu$, the one- and two-loop contributions become of the same order of magnitude for slepton masses around 3-10 TeV. To determine the
values of the Higgs, gaugino, and slepton mass parameters for which
\mbox{$d_e$(2-loop)$> d_e$(1-loop)} one may use Figs.~2,~3. For
example, taking $\sin\phi_\mu=0.1$ we observe from Fig.~3 that the
one-loop contribution to $d_e$ becomes smaller than the two-loop
contribution for $m_{\tilde\ell}\gsim 2$ TeV for $|\mu|, M_1 (M_2) \sim
100 $ GeV for the GUT (AMSB) gaugino mass hierarchy. From the plots in
Fig.~2{\em(c,d)} we find that all values of $|\mu|, M_{1,2}$ lying to the
lower left of the dark blue line lead to to $d_e$(2-loop) that is
larger than the experimental limit for $\sin\phi_\mu=0.1$. In this case,
only the portions of the funnel lying to the upper right of the dark blue
curve are compatible with the two-loop EDM limits and the observed BAU.

The constraints for different values of $\sin\phi_\mu$ or for different
experimental limits may be obtained by rescaling the curves (horizontal
lines) in Fig.~2 (Fig.~3) accordingly. Again to illustrate, for
$\sin\phi_\mu=0.5$ we obtain from Fig.~3 that \mbox{$d_e$(2-loop}$>d_e$(1-loop) for $m_{\tilde\ell}\gsim$4-5 TeV. The
corresponding curves in Fig.~2{\em(b,c)} can be obtained by dividing the
current limit on $d_e$ by $\sim 5$, leading to the curves labeled
$d_e=4\times 10^{-28}$ e-cm. Note that for larger values of
$\sin\phi_\mu$, the reach of the EWB compatible funnels extends further
out (see Fig.~1), but in this case ($\sin\phi_\mu=0.5$, $m_A= 150$ GeV),
they fall entirely within the range of Higgsino and gaugino mass
parameters incompatible with the two-loop EDM limits.

To summarize, we find that (i) the smallest CPV phase compatible with EWB in the present setting is $\sin\phi_\mu\simeq10^{-2}$, (ii) the current two-loop electron EDM constraints rule out MSSM EWB with maximal CPV phase, (iii) even with superheavy sleptons, future electron EDM experiments feature a sensitivity which will largely cover all the viable MSSM EWB parameter space and (iv) two-loop contributions to $d_e$ dominate one-loop contributions for slepton masses larger than a few TeV, depending on the values of the heavy Higgs masses.

\section{Dark Matter Implications}\label{sec:dm}

Resonant EWB implies that one of the gaugino masses has to be almost degenerate with the higgsino mass parameter $\mu$ (see the discussion in Sec.~\ref{sec:mssm} and Fig.~\ref{fig:yb}). This entails that either a bino-like or  wino-like neutralino features a significant degree of gaugino-higgsino mixing. This mixing, in turn, can have dramatic consequences for the phenomenology of the lightest supersymmetric particle --  particularly if the mixed state coincides with the lightest neutralino.

\begin{figure}[!t]
\begin{center}
\epsfig{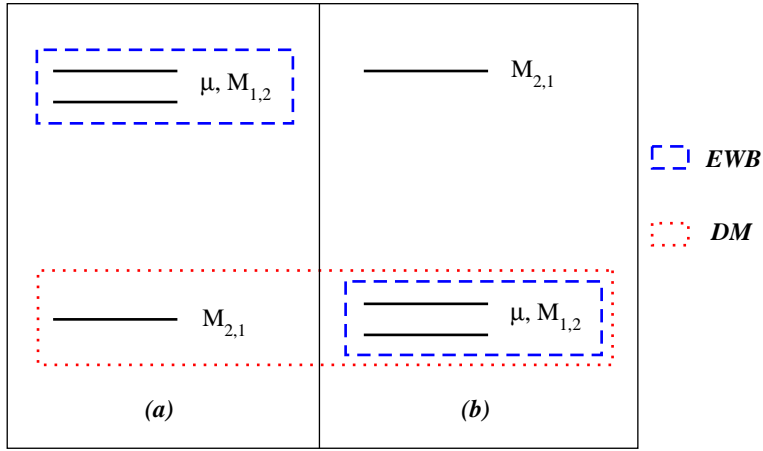}
\end{center}
\caption{\it\small A sketch of the possible spectral patterns for the parameters entering the neutralino mass matrix (the higgsino mass parameter $\mu$ and the gaugino soft breaking masses $M_{1,2}$), in the context of resonant EWB. In case $(a)$ the sector relevant for dark matter phenomenology is decoupled from the one responsible for EWB, while in case $(b)$ the two sectors are connected: the same fields which drive the resonant EWB non-trivially participate in the determination of the dark matter particle mass and composition.
}
\label{fig:sketch}
\end{figure}

We give a pictorial sketch of the possible neutralino mass matrix parameters hierarchies giving rise to resonant EWB in Fig.~\ref{fig:sketch}: the DM sector (red dotted lines) can be decoupled from the sector responsible for resonant EWB (case {\em (a)}), or the two sectors can overlap (case {\em (b)}), giving rise to the interesting possibility of a close DM-EWB connection. The two possibilities clearly emerge also from a closer look to Fig.~\ref{fig:yb}, where the portions of the ($M_{1,2},\mu$) parameter space compatible with the BAU split in two branches: in the upper branch the pattern of the neutralino mass matrix parameters is given by case {\em (a)} in Fig.~\ref{fig:sketch}, while in the lower branch the parameters are arranged as in case {\em (b)}.

In what follows we discuss the DM phenomenology of MSSM EWB-compatible models. We start in Subsec.~\ref{sec:omega} by analyzing the requirements stemming from the thermal relic abundance of neutralinos. In Subsec.~\ref{sec:dmsearches} we outline the prospects of detection of dark matter at direct and indirect search experiments, focusing in particular on the detection of the neutralino-induced energetic neutrino flux from the Sun, and comparing the detection rates in the case of maximal and null CP violation.  Lastly, we devote Subsec.~\ref{sec:colliders} to an overview of current and future collider searches of MSSM models compatible with EWB.

\subsection{Dark Matter relic abundance}\label{sec:omega}

In this Section we assess the consistency of a thermal relic abundance, as predicted in a standard neutralino decoupling occurring in the radiation-dominated era of the early universe, with the BAU generated through EWB. We compute the relic abundance of neutralinos with a specially customized version of the {\tt DarkSUSY} package \cite{ds} that takes into account the effects of the non-vanishing CP-violating phase $\phi_\mu$. Regions of parameter space with a thermal relic abundance larger than the 2-$\sigma$ cold dark matter abundance determined by the WMAP collaboration in the context of a $\Lambda$CDM cosmology are strongly phenomenologically disfavored. The only viable option for those models would be to dilute the abundance of neutralinos through late entropy injection, {\em i.e.} through a late ``reheating''. This mechanism, however, is highly constrained by the requirement of preserving the successful predictions of the primordial nucleosynthesis of light elements (for a different point of view, see also Ref.~\cite{Gelmini:2006pw}). 

If, in contrast, the thermal production of neutralinos in the early universe is insufficient to provide all of the cold dark matter content of the universe, various options make those low thermal relic abundance models phenomenologically viable. For example,  neutralinos could constitute just a fraction of the cold dark matter, with other particles making up for the rest; in this case, detection rates should be properly rescaled to account for the fraction of dark matter effectively ascribed to the LSPs. Alternately, a modified cosmological history of the early universe can also lead to different values of the thermal relic abundance itself: for instance, if the energy density of the universe at the time of neutralino decoupling was dominated by an extra component, be it a quintessential dark energy scalar field \cite{quint} or the effective energy density associated to a primordial anisotropic shear \cite{cosmoen}, or a Brans-Dicke-Jordan modified cosmological expansion rate \cite{bdj}, the Hubble parameter would have been larger, forcing an earlier freeze-out of neutralinos. This, in turn, entails a larger thermal relic abundance, with enhancement effects in principle as large as six orders of magnitude \cite{cosmoen}. A third option is the non-thermal production of neutralinos through the decays of metastable species, with lifetimes larger than the time when neutralinos decoupled from the thermal bath in the early universe \cite{nonth}. 

Here, we will assume that one of the two latter options brought low relic abundance models in accord with a scenario where all the CDM is made of neutralinos. We will not, therefore, proceed with any rescaling of the detection rates in the case of low thermal relic abundance models.
Given a thermal relic abundance $\Omega_\chi h^2<\Omega_{\rm CDM} h^2$, where $\Omega_{\rm CDM}h^2\simeq0.11$, we define a parameter
\begin{equation}
\eta_\Omega\equiv\frac{\Omega_{\rm CDM}-\Omega_\chi}{\Omega_\chi}
\end{equation}
which quantifies the  relative enhancement factor needed to bring a low relic abundance model into accord with the CDM abundance. Equivalently, $\eta_\Omega$ can be regarded as the relative number of non-thermally versus thermally produced neutralinos: given a low relic abundance model, and assuming a mechanism of non-thermal production, a model associated to a certain value of $\eta_\Omega$ needs $\eta_\Omega$ non-thermally produced neutralinos per thermally produced neutralino. 

\begin{figure}[!t]
\begin{center}
\hspace*{-1.cm}\epsfig{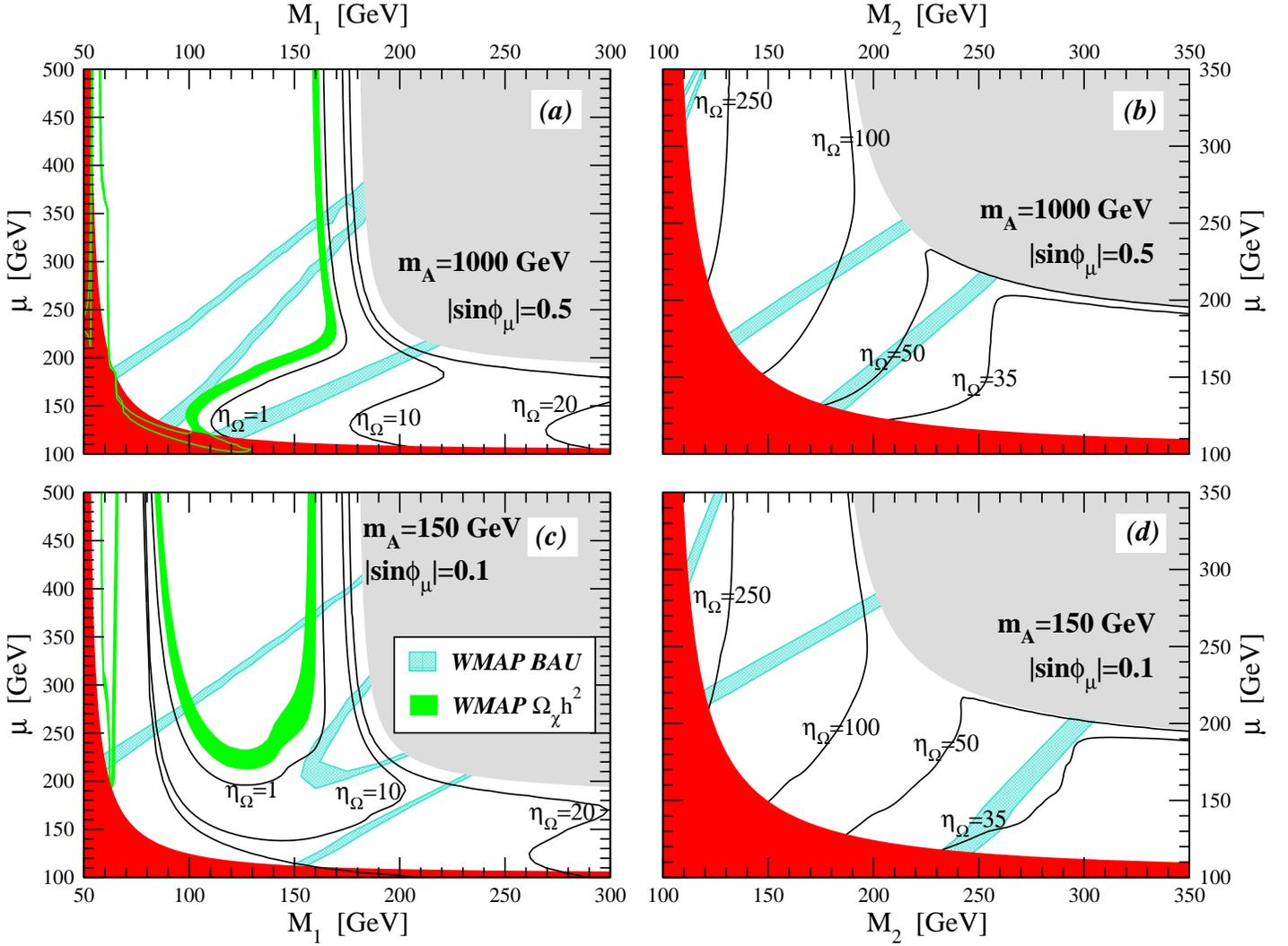}
\end{center}
\caption{\it\small 
A close-up on the relic abundance of neutralino dark matter on the $(M_{1,2},\mu)$ planes. The red regions are excluded by the LEP2 chargino mass bound, while in the grey regions the stop becomes the LSP. The conventions for the values of $m_A$, $\sin\phi_\mu$ and for the gaugino soft breaking masses are as in the previous Fig.~\protect{\ref{fig:edm}}. We shade in light green regions of the parameter space where the neutralino thermal relic abundance falls within 2-$\sigma$ in the CDM abundance range determined by WMAP \cite{Spergel:2003cb}. We indicate with black curves the contours of constant $\eta_\Omega$, the number of ``extra'' neutralinos needed per thermally produced neutralino to bring low thermal relic abundance models into accord with the inferred CDM density.}
\label{fig:oh2}
\end{figure}
In this Section we study the regions in the $(M_{1,2},\mu)$ planes compatible with the EWB generation of the BAU and with the relic abundance of neutralinos, including the possibility of a primordial enhancement, parameterized with different values of $\eta_\Omega$. In Fig.~\ref{fig:oh2} we study a close-up of the small mass region of the planes considered back in Fig.~\ref{fig:edm}, {\em i.e.} the ($M_1,\mu$) (panels {\em (a)} and {\em (c)}) and ($M_2,\mu$) (panels {\em (b)} and {\em (d)}) planes, at $m_A=1000$ ({\em (a)} and {\em (b)}) and 150 GeV ({\em (c)} and {\em (d)}). We shade in red and grey, respectively, the regions incompatible with the experimental bounds on the mass of the lightest chargino \cite{lep2}, and where the LSP is no longer the lightest neutralino, but rather the lightest, right-handed stop (which, we recall, is assumed to have a mass close to the top mass). We shade in light blue the regions producing a BAU within the 2-$\sigma$ WMAP range \cite{Spergel:2003cb} with $|\sin\phi_\mu|=0.5$ ({\em (a)} and {\em (b)}) and 0.1 ({\em (c)} and {\em (d)}). Finally, regions giving rise to a thermal neutralino relic abundance within the WMAP 2-$\sigma$ range \cite{Spergel:2003cb} are indicated with a green shading. 

Panel {\em (a)} in Fig.~\ref{fig:oh2} illustrates the various possible scenarios where neutralinos end up with the correct thermal relic abundance in a standard cosmological setup. First, we point out that all these cosmologically allowed regions lie in the portion of parameter space where $M_1<\mu$ and the neutralino is (almost) bino-like: since \mbox{$m_\chi<m_t$} and $m_f\ll m_\chi$ for other fermions, $t$- (or $u$-)channel annihilations of binos into SM fermions ($\chi\chi\stackrel{t,u(\widetilde f)}{\longrightarrow}\bar f f$) are suppressed, implying inefficient pair annihilations of the LSP in the early universe\footnote{We recall that the lightest stop does not contribute to this class of diagrams, since the top final state is here kinematically forbidden; even if other sfermions were lighter than what we consider here, the $s$-wave suppression factor $(m_f/m_\chi)^2$ of a final fermion-antifermion state would still make these further contributions too small to significantly affect the bino relic abundance.}. When supplementary mechanisms participate in suppressing the bino relic abundance, the density of relic neutralinos can give the right CDM density inferred from cosmology, unlike the other parameter space regions where the relic abundance is systematically below the CDM abundance. In panels {\em (a)} and {\em (c)}, regions lying above the green shaded strips are excluded because of overproduction of thermal relics. 

As alluded, though, various mechanisms can conspire to bring, in special regions of the parameter space, the neutralino relic abundance in accord with the CDM abundance. A first region corresponds to the thin, almost vertical strip on the left in panels {\em (a)} and{\em (c)}, produced by the resonant annihilation of almost purely bino-like neutralinos ($M_1\ll\mu$) through the lightest Higgs boson $h$, when $m_\chi\simeq m_h/2$, in the annihilation mode $\chi\chi\stackrel{s(h)}{\longrightarrow}\bar f f$. The second cosmologically favored region lies in the center of the plot, and features a mixed bino-higgsino LSP ($M_1\simeq\mu$). This configuration implies on the one hand sizable neutralino annihilation rates into gauge bosons, {\em e.g.} through the reaction $\chi\chi\stackrel{t,u(\widetilde\chi^\pm_i)}{\longrightarrow}W^+W^-$, and, on the other hand, coannihilation processes of the LSP with the (higgsino-like) next-to-lightest neutralinos and with the lightest chargino,{\em e.g.} $\chi\widetilde\chi^\pm\stackrel{t,u(\widetilde \chi^\pm_i,\widetilde\chi^0_j)}{\longrightarrow}W^\pm Z$, similarly to what happens in the focus point region of mSUGRA \cite{fp}. Unlike the focus point region, however, in the present scenario when the neutralino mass gets closer to the stop mass, coannihilations with the lightest stop \cite{stop} ({\em e.g.} $\chi\widetilde t_1\stackrel{s(t)}{\longrightarrow}b W$) contribute significantly to suppress the neutralino relic abundance down to the level of the CDM abundance. This third region corresponds to the almost vertical strips in panels {\em (a)} and{\em (c)} lying close to the grey region where the lightest stop is lighter than the neutralino $\chi$. Lastly, in panel {\em (c)}, a region of parameter space at $m_\chi\simeq m_A/2=75$ GeV also becomes viable at large $\mu$ and $50\lesssim M_1/{\rm GeV}\lesssim100$ by virtue of resonant annihilations with the heavy Higgses ($\chi\chi\stackrel{s(A,H)}{\longrightarrow}\bar f f$).

From panels {\em (a)} and{\em (c)} we deduce that, depending on $m_A$ and on the CPV phases, in the $M_1\ll M_2$ case various regions of the MSSM parameter space can (1) thermally produce the correct neutralino relic abundance, and (2) produce the appropriate BAU through resonant MSSM EWB. For instance, at large $m_A$, {\em (a)}, one can satisfy both requirements in the $h$ resonant annihilation region, in the mixed bino-higgsino region and in the stop coannihilation region; at small $m_A$, {\em (b)}, the two conditions are fulfilled again in the $h$ resonant annihilation region and in the stop coannihilation region, as well as in the parameter space portions corresponding to regions of resonant annihilations through heavy Higgses as well. If $M_2\ll M_1$, instead, the thermal relic abundance of neutralinos is always significantly smaller than the CDM abundance, by virtue of multiple neutralino/chargino coannihilations and efficient annihilation rates into gauge bosons. In all cases, the CDM-allowed and BAU-allowed regions corresponds to wino-higgsino driven baryogenesis, whereas the regions corresponding to bino-higgsino (pure neutralino) EWB lead to insufficiently large relic densities.

As outlined above, however, even if the thermal relic abundance of neutralinos lies below the CDM abundance, non-thermal mechanisms or modified cosmological scenarios can enhance the final population of neutralinos. It is illustrative, here, to point out  the preferred ranges of such enhancements, which can, in principle, provide an indication on the nature of the involved non-thermal processes and constrain the relative phenomenology. To this end, we indicate with black lines the curves at constant values of the enhancement factor $\eta_\Omega$. Increasing the higgsino nature of the LSP ({\em i.e.} moving to the lower right portions of each panel) suppresses the LSP relic abundance if $M_1\ll M_2$, and the required enhancement factors can be as large as $\approx 30$. The iso-level curves then tend to flatten over the $m_\chi\simeq m_{\widetilde t_1}$ line ({\em i.e.} the grey contour), again due to stop coannihilations. Instead, if $M_2\ll M_1$ (panels to the right), an increasing higgsino fraction enhances the final LSP relic abundance, since a purely wino-like neutralino/chargino system features larger effective annihilation rates and a reduced number of coannihilating degrees of freedom with respect to a mixed or a purely higgsino-like system. In this case, the range of enhancement factors is between 30 and 300. This means, for instance, that consistency with EWB requires -- in a standard cosmological scenario -- that for every thermally produced neutralino at least 30 to 300 neutralinos must be non-thermally produced.

\subsection{Dark Matter searches}\label{sec:dmsearches}
\begin{figure}[!t]
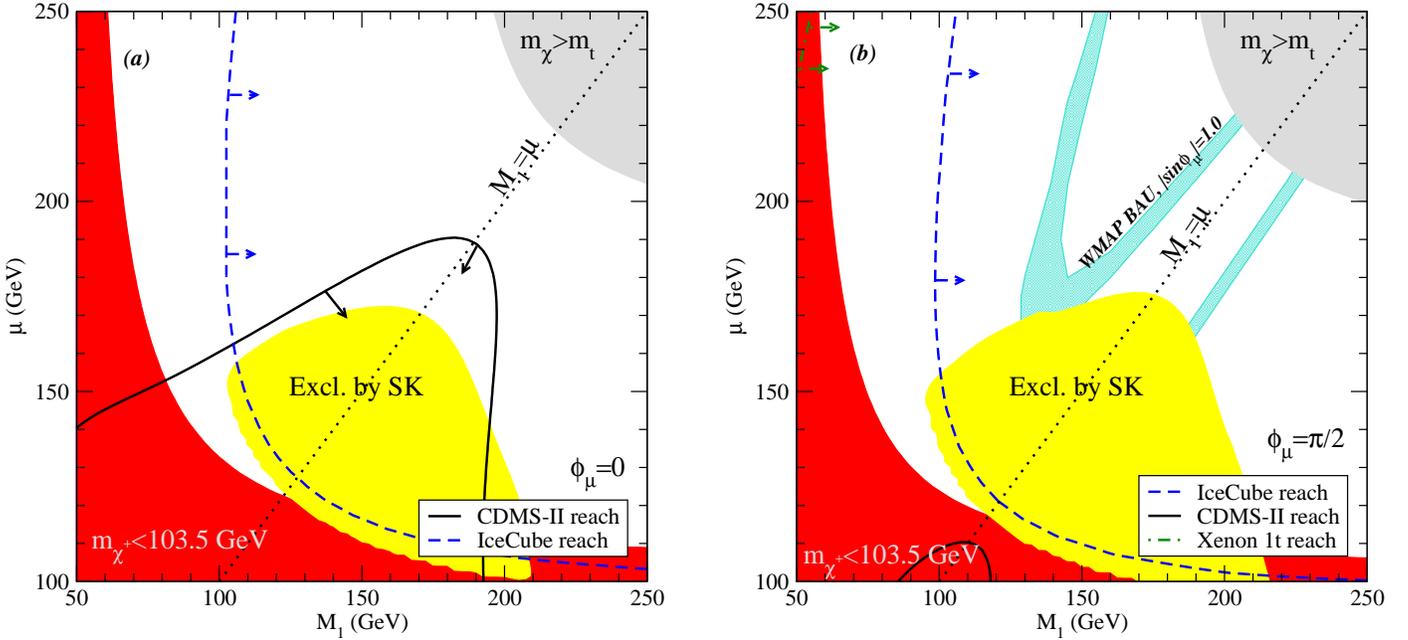

\begin{center}
\mbox{\hspace*{-0.8cm}\epsfig{file=plots/m1mu_0.eps,height=8.5cm}\qquad\epsfig{file=plots/m1mu_1.eps,height=8.5cm}}
\end{center}
\caption{\it\small Dark matter detection rates without (a) and with maximal (b) CP violation in the higgsino sector, on the $(M_1,\mu)$ plane, at $m_A=$1 TeV. Red and grey regions indicate parameter space portions where the LEP2 bound on the chargino mass is violated and where the neutralino is not the LSP, respectively. The yellow region is excluded by the SuperKamiokande data on the flux of energetic neutrinos from the Sun \cite{Habig:2001ei}, while the regions lying to the right of the blue dashed lines will produce a sizable flux of neutrinos at IceCube \cite{icecube}. The reach of CDMS-II \cite{cdms2} is indicated by a black solid line, while all the viable parameter space shown in the two panels will be within reach of next generation ton-sized direct detectors, such as Xenon \cite{xenon}. The light blue shaded region corresponds to values of the supersymmetric parameters which generate the desired BAU, according to \cite{Spergel:2003cb}.}
\label{fig:m1mu}
\end{figure}

The purpose of this Section is two-fold: First, we want to determine if the regions of the MSSM parameter space compatible with EWB will produce sizable signals at future dark matter search experiments, and if current data already rule out portions of the parameter space; Second, we wish to study how dark matter detection rates --  particularly indirect detection channels -- are affected by the occurrence of non-trivial CP-violating phases.

\begin{figure}[!t]
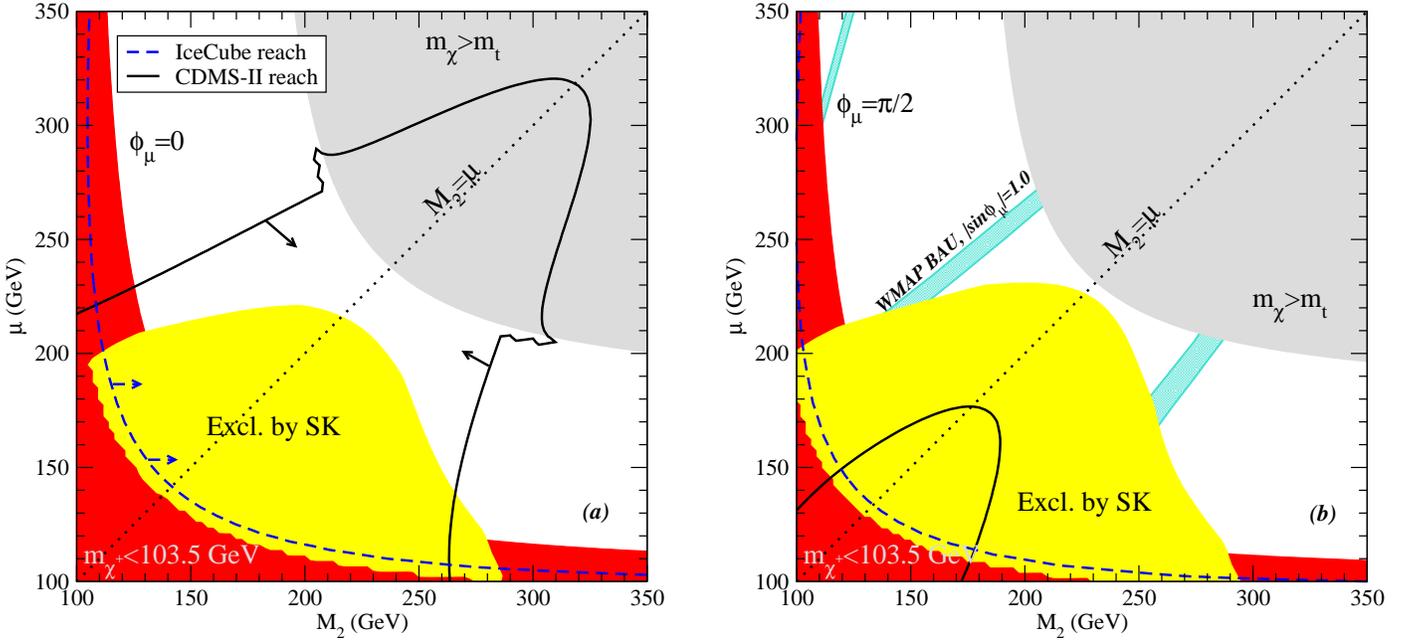

\begin{center}
\mbox{\hspace*{-0.8cm}\epsfig{file=plots/m2mu_0.eps,height=8.5cm}\qquad\epsfig{file=plots/m2mu_1.eps,height=8.5cm}}
\end{center}
\caption{\it\small Same as in Fig.~\protect{\ref{fig:m1mu}}, but on the $(M_2,\mu)$ plane.}
\label{fig:m2mu}
\end{figure}

To this end, we compare in Fig.~\ref{fig:m1mu} the dark matter sensitivity reach in direct and indirect detection experiments for the CP conserving {\em (a))} and maximally CP violating case (\mbox{$|\sin\phi_\mu|=1$}, {\em (b)}), in the $(M_1,\mu)$ plane, with GUT-scale gaugino mass unification. We resort here to the extreme case \mbox{$|\sin\phi_\mu|=1$} to illustrate the maximal possible effect induced by CP violation on the direct and indirect DM detection rates. Also, we pick $m_A=$1 TeV (smaller values of the heavy Higgs mass scale would imply, in general, larger rates). In Fig.~\ref{fig:m2mu}, instead, we consider an anomaly-mediated gaugino mass spectrum, and study the $(M_2,\mu)$ plane. As in the previous figures, red shading indicates a chargino below the direct accelerator search limits, and grey shading implies $m_\chi>m_{\widetilde t_1}$. In the panel to the right, we also indicate with a light blue shading the parameter space region compatible with the WMAP 2-$\sigma$ BAU.

We consider here the planned sensitivity of forthcoming Stage 2 detectors (namely, the next stage of the CDMS-II experiment, \cite{cdms2}) and that of ton-sized, Stage 3 detectors (for definiteness we use the planned sensitivity of the Xenon 1-t experiment, \cite{xenon}). As far as indirect detection is concerned, we concentrate on neutrino telescopes detection of the muons produced by charged current interactions of neutrinos generated by annihilations of neutralinos gravitationally trapped in the core of the Sun. This channel features a very mild dependence on the details of the dark matter halo model, and, unlike antimatter searches, the rates do not depend on the (model-dependent) diffusion mechanisms of charged particles in the galaxy. In particular, we study the impact of the results of the SuperKamiokande collaboration on the flux of energetic neutrinos from the Sun, Ref.~\cite{Habig:2001ei}. We also estimate the reach of future ${\rm km}^3$ neutrino telescopes, such as IceCube \cite{icecube}, enforcing a muon energy threshold of 50 GeV, and requiring a sizable signal flux, namely 10 events per ${\rm km}^2$ per year. The reach contours are only mildly affected when assuming a larger energy threshold and a lower signal flux.

In Figs.~\ref{fig:m1mu} and \ref{fig:m2mu} we indicate with a black solid line the reach of the next stage CDMS-II experiment: points inside the contours are predicted to give a detectable signal. The sensitivity of ton-sized direct detectors will cover the entire panel {\em (a)} in Fig.~\ref{fig:m1mu}, almost all panel {\em (b)} (but a tiny region in the top left, already ruled out by chargino searches) and both panels of Fig.~\ref{fig:m2mu} %

The yellow shaded area is ruled out by the SuperKamiokande data. This is one of the main results of the present analysis: a significant portion of the MSSM parameter space compatible with EWB is already ruled out by the SuperKamiokande data. The constraints are particularly stringent in the regions where resonant baryogenesis involves the LSP sector, as we outlined above (case {\em (b)} in Fig.~\ref{fig:sketch}). The $\mu\sim M_{1,2}$ condition maximizes the scalar neutralino scattering cross section off nuclei, proportional to the product of the LSP gaugino and higgsino fractions. In this latter case, the BAU-viable parameter space is contained  within the two lower branches of correct BAU surrounding the resonant $M_{1,2}\sim\mu$ line (lowering $|\sin\phi_\mu|$ one is forced to lie closer to the resonance, see Fig.~\ref{fig:yb}). In particular, in the $M_2\ll M_1$ case almost all of the $M_2\simeq\mu$ region is ruled out by the SuperKamiokande bound, enforcing either $m_\chi\approx m_{\widetilde t_1}\approx m_t$ and $M_2\simeq\mu$, or close-to-maximal CP violating phases, and off-resonance values of $M_2$ and $\mu$. A third possibility is that of neutralino EWB in the upper funnel at $\mu\simeq M_1$ (upper left corner of the figure). As far as the IceCube reach is concerned, all the EWB-compatible parameter space shown in the two figures will give a sizable signal, predicted to be well above the level needed to disentangle it from the background.

\begin{figure}[!t]
\begin{center}
\hspace*{-1.cm}\epsfig{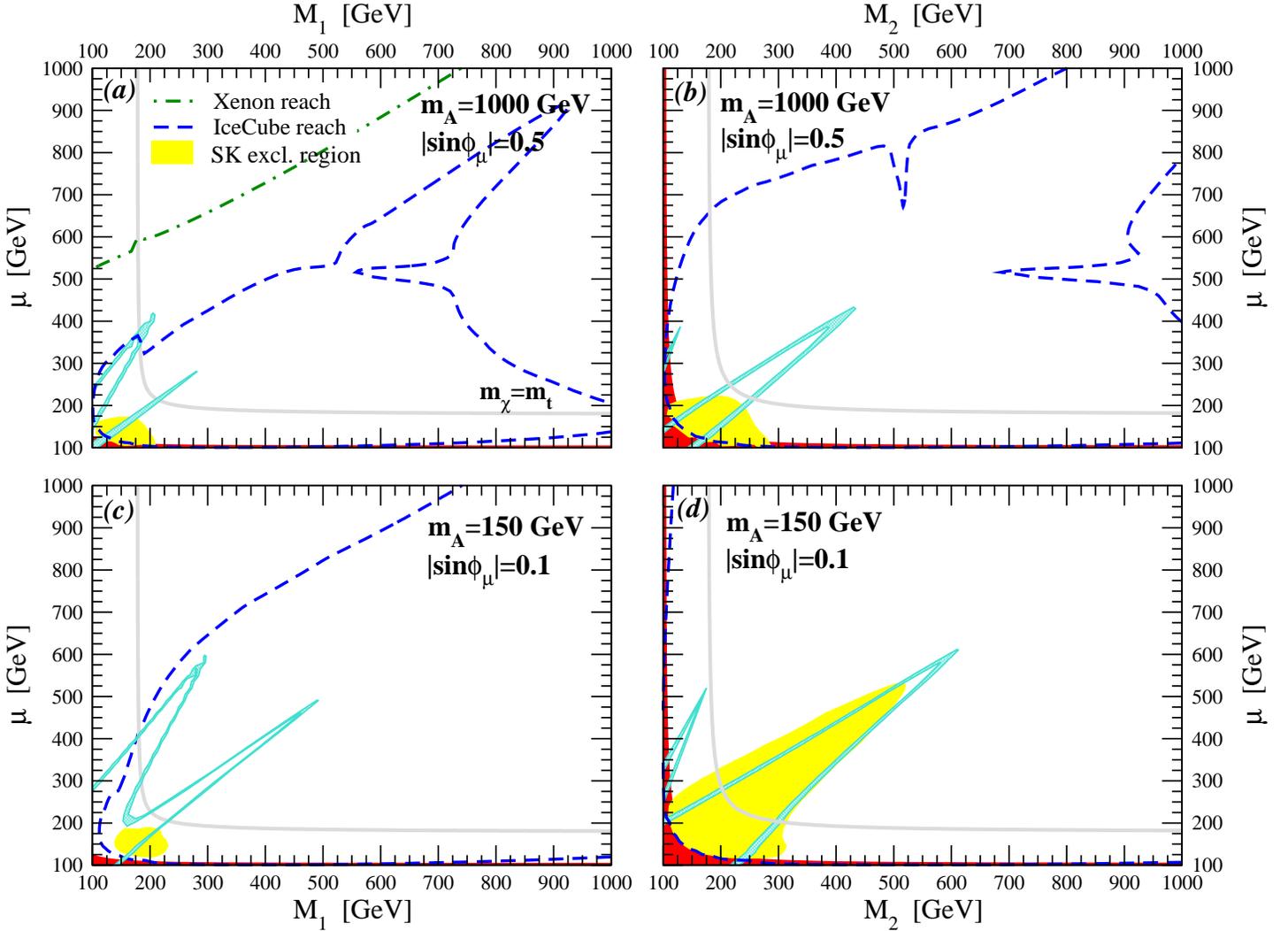}
\end{center}
\caption{\it\small Current exclusion limits from the SuperKamiokande collaboration \cite{Habig:2001ei} (yellow shaded region) and future sensitivity reach of IceCube \cite{icecube} and of Xenon-1t \cite{xenon} on the $(M_{1,2},\mu)$ planes at $|\sin\phi_\mu|=0.5$ (panels {\em (a)} and {\em (b)}) and 0.1 (panels {\em (c)} and {\em (d)}); we also show the contours of maximal neutralino masses compatible with a stop as heavy as the top quark (grey lines), and the regions of the parameter space which produce a BAU compatible, at the 2-$\sigma$ level, with the WMAP result \cite{Spergel:2003cb}. The two upper panels refer to a heavy scalar mass $m_A=1000\ {\rm GeV}$, while the two lower panels employ $m_A=150\ {\rm GeV}$.}
\label{fig:nt}
\end{figure}
We study in greater detail the IceCube reach and the Xenon-1t reach in Fig.~\ref{fig:nt}, where we extend the axis ranges to larger values, in order to understand the size of the largest possible neutralino masses that could be detected by the two forthcoming experiments with maximal CP violating phase. Assuming alternative mechanisms are operative to render the electro-weak phase-transition more strongly first order than in the MSSM \cite{nonminimal}, one can relax the assumption of having a light right-handed stop, and hence the inferred limit $m_\chi\lesssim m_t$ on the neutralino mass (which we still indicate for reference in Fig.~\ref{fig:nt} with a grey line). In Fig.~\ref{fig:nt} we also use lower values of the heavy Higgs mass $m_A$, and reproduce the 2-$\sigma$ WMAP contours for the BAU, at $\sin\phi_\mu=0.5$ and 0.1 in the upper and lower panels respectively. We still assume here that even within a next-to-minimal setup where the lighter stop does not need to be light, the main source of CP violation comes from the gaugino/higgsino sector, and not from other sources, plausible in this more general context ({\em e.g.} resonant effects in the squark sector). As before, the yellow shaded area is excluded by the SuperKamiokande data, and we use, for the assessment of the IceCube sensitivity, a muon energy threshold of 50 GeV and a minimal detectable flux of 10 events per ${\rm km}^2$ per year. We stress again how the SuperKamiokande results on energetic neutrinos from the Sun strongly constrain the supersymmetric parameter space compatible with resonant EWB BAU (see in particular panel {\em (d)}).

We notice that in the pure higgsino limit ($\mu\ll M_{1,2}$), where the ${\tilde\chi}^0{\tilde\chi}^0 Z$ coupling is large, and where the spin-dependent neutralino-nucleon scattering cross section is therefore sizable, the IceCube reach extends to neutralino masses even heavier than 1 TeV. In addition,  for $2\cdot m_\chi\simeq m_A=$ 1000 GeV the neutrino telescope rates are suppressed, even though the pair annihilation cross section is greatly enhanced by the heavy Higgses resonances. This situation follows from the fact that over most of the parameter space we consider, neutralino annihilations and captures in the core of the Sun are in equilibrium: an increase in the pair annihilation cross section does not lead to an increase in the annihilation rate inside the Sun, which in equilibrium only depends upon the capture rate. Rather, on top of the resonance, the branching fraction into neutrinos is suppressed with respect to the off-resonance case, therefore yielding a reduced neutrino flux and the consequent suppression of the neutrino telescope rates. The Xenon-1t reach covers essentially all of the BAU-compatible parameter space; in the planes shown in Fig.~\ref{fig:nt}, only the purely bino-like regions (where the neutralino-neutralino-Higgs couplings are suppressed), at large $m_A$, will be beyond the sensitivity of ton-size detectors\footnote{We stress, though, that those regions will hardly produce the right BAU through resonant MSSM EWB.}: panels {\em (b)}, {\em (c)} and {\em (d)} will be fully within the Xenon-1t reach. 

Comparing the IceCube and the Xenon-1t reach contours with the regions consistent with EWB, we infer that, in the ($M_2,\mu$) plane, (panels {\em (b)} and {\em (d)}),  the IceCube and the Xenon-1t future sensitivities extend over the whole BAU-compatible parameter space, both in the EWB-DM connection branch and in the EWB-DM disconnected branch. In the ($M_1,\mu$) plane, panels {\em (a)} and {\em (c)}, instead, IceCube will be sensitive to the whole EWB-DM connection branch, while the $M_2\sim\mu$ regions, with a mostly bino-like LSP, will likely give sizable signals, but, in general, may be beyond reach. However, those regions will be fully explored by forthcoming ton-sized direct detection experiments. We therefore conclude that  under general assumptions about the MSSM parameter space, neutrino telescopes will  be a powerful probe of supersymmetric dark matter if electro-weak baryogenesis is the mechanism of generation of the BAU, unlike the generic situation in the MSSM, where neutrino fluxes from neutralino annihilations in the Sun can be significantly suppressed \cite{Profumo:2004at}. Further, future direct detection experiments will be able to probe most, if not all, of the resonant EWB-compatible parameter space, even if the LSPs do not participate directly in the dominant baryogenesis mechanism.

\begin{figure}[!t]
\begin{center}
\epsfig{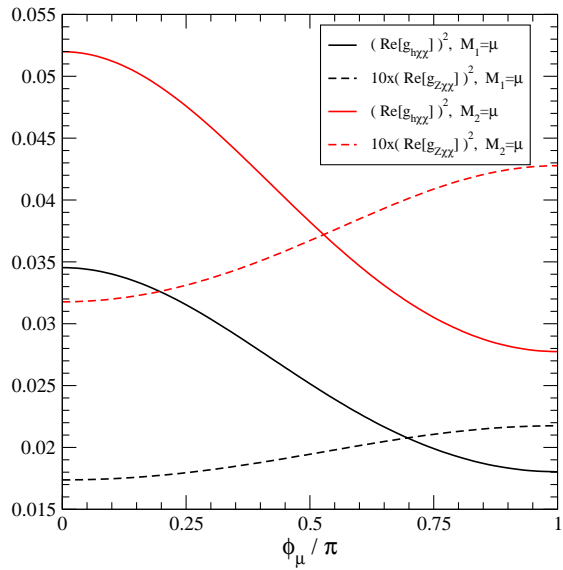}
\end{center}
\caption{\it\small 
The variation of the neutralino-neutralino-Higgs/$Z$ boson couplings squared, respectively relevant for the scalar and axial scattering rate of neutralinos off matter, for $M_1=\mu=200$ GeV, $M_2=400$ GeV (black lines) and $M_2=\mu=200$ GeV, $M_1=600$ GeV (red lines).}
\label{fig:coup}
\end{figure}
A conspicuous feature of the direct versus indirect detection reach contours shown in Fig.~\ref{fig:m1mu} and \ref{fig:m2mu} is that increasing the CP violating phase $\phi_\mu$ from 0 to $\pi/2$ leads to smaller direct detection rates and larger rates at neutrino telescopes. In order to quantify this statement and to understand the underlying physical reasons, we assessed that the largest contributions to the spin-independent and spin-dependent neutralino-nucleon scattering cross sections that respectively correspond, at large $m_A$, to the $t$-channel lightest Higgs $h$ and $Z$ exchange. In turn, the most important diagrams will depend upon the square of the $\chi$-$\chi$-$h$ and $\chi$-$\chi$-$Z$ couplings. We study the variation of these two quantities with $\phi_\mu$ in Fig.~\ref{fig:coup} for two representative MSSM parameter space points, the first at $M_1=\mu=200$ GeV, $M_2=400$ GeV (black lines) and $M_2=\mu=200$ GeV, $M_1=600$ GeV (red lines)\footnote{We multiplied the $\chi$-$\chi$-$Z$ couplings squared by a factor 10 for illustrative purposes.}. The net effect of increasing the phase $\phi_\mu$ is therefore to enhance the spin-dependent neutralino-nucleon scattering cross section, and to suppress the spin-independent cross section, as expected. We also notice that the case of maximal CP violating phases, $\sin\phi_\mu=1$, interpolates, in both cases, between the rates expected, in the CP conserving case, with $\mu>0$ and $\mu<0$. 

\subsection{Collider searches: overview}\label{sec:colliders}

\begin{figure}[!t]
\begin{center}
\epsfig{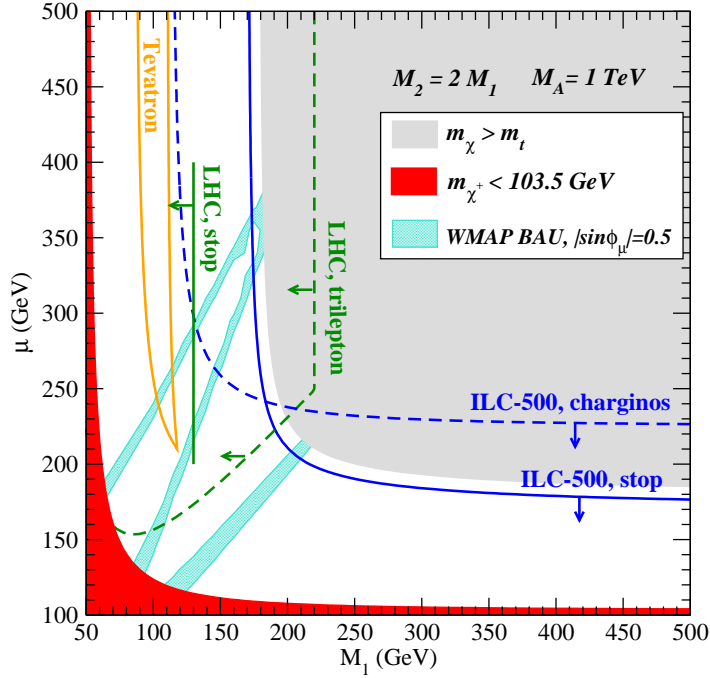}
\end{center}
\caption{\it\small A qualitative overview of next generation colliders reaches on the $(M_1,\mu)$ plane, at $m_A=1000$ GeV and $\sin\phi_\mu=0.5$. The color code for the shadings is the same as in the previous figures, and details on the sensitivity lines are given in the text.}
\label{fig:colliders}
\end{figure}

In addition to EDM and DM searches, present and future collider experiments will probe the parameter space relevant to resonant MSSM EWB. Here, we summarize the corresponding sensitivity of studies at the Tevatron, LHC, and ILC\cite{Allanach:2006fy}. In Fig.~\ref{fig:colliders} we sketch the plausible reach contours for the Tevatron, the LHC, and for a 0.5 TeV center-of-mass energy $e^+e^-$ linear collider, assuming GUT-scale gaugino unification ($M_2\simeq 2\cdot M_1$), a light right-handed stop, $m_{\widetilde t_1}\simeq m_t$, and $m_A=1$ TeV. The red and grey shadings are as in the previous figures. The arrows point to the regions of parameter space which will be within reach.

For the Tevatron, we assume an integrated luminosity of 4 ${\rm fb}^{-1}$. It was shown in Ref.~\cite{Demina:1999ty} that light stops can be optimally searched for in two stop decay channels: 

\begin{itemize}

\item[i)] If the stop is heavier than the lightest chargino, the dominant stop decay is $\widetilde t_1\rightarrow b\widetilde\chi^+_1$, and the most promising signature is a $b$-jet, an isolated lepton, a jet and missing transverse energy \cite{Demina:1999ty}. In this case, however, the reach of Tevatron with 4 ${\rm fb}^{-1}$ of integrated luminosity only extends up to a chargino mass of around 100 GeV, for stop masses of interest here, and therefore the resulting parameter space sensitivity only covers models which are already ruled out by the LEP2 bound on the chargino mass. 

\item[ii)] On the other hand, if the stop is lighter than the lightest chargino, then the stop will dominantly decay into the LSP and a $c$-quark (either trough loop-suppressed diagrams or at tree level through ${\tilde t}$-${\tilde c}$ mixing). In this case, the expected signal is two acolinear charm jets and missing transverse energy. The corresponding Tevatron reach, at stop masses close to the top mass extends to neutralino masses as large as 110 GeV, provided $m_{\widetilde t_1}-m_{\widetilde\chi}\gtrsim30$ GeV. 

\end{itemize}

In Fig.~\ref{fig:colliders}, the Tevatron reach contour is bounded to the left by the requirement that $m_{\widetilde t_1}>m_{\widetilde\chi^+_1}$, and to the right by the requirement of a sufficiently light neutralino. Assuming the mass of lightest stop to be close to the top quark mass, the neutralino-stop mass difference constraint is always fulfilled. The Tevatron sensitivity is therefore confined here to a bino-like lightest neutralino, which can be -- in some models -- compatible with a WMAP neutralino thermal relic abundance, and which also partly overlaps the $M_2\sim\mu$ resonant EWB funnel.

At the LHC, limitations due to background levels and detector thresholds will make searches for a light stop extremely hard, even though the expected number of signal events is by far larger than at the Tevatron. Although exhaustive dedicated analysis has not yet been carried out, a promising signature of a light-stop scenario at the LHC appears to be that of same-sign top quarks \cite{Kraml:2005kb} (see also \cite{Allanach:2006fy}). If gluinos are light enough to be abundantly produced, they will prominently decay into top-stop pairs; pair-produced gluinos will then give rise, in almost half the cases, to same-sign top events ($\widetilde g\widetilde g\rightarrow\widetilde t_1^* t\ \widetilde t_1^* t$). Letting the top quarks decay into $bW$ (and the $W$ subsequently decaying leptonically) and the stops into $c\chi$, then leads to a significantly distinctive signature: two same-sign $b$ quarks, two same sign leptons, jets and missing transverse energy. However, if the chargino is lighter than the stop, the latter will dominantly decay into $b\widetilde\chi^+_1$ rather than $c\chi$, losing the same-sign top quarks information. In Fig.~\ref{fig:colliders} we give the LHC reach in this channel labeling it as ``LHC, stop'', and referring to the largest possible gluino mass which this channel can probe according to Ref.~\cite{Kraml:2005kb}. In the region at smaller $\mu$ below the lower end of the line, and at values of $M_1\lesssim 80$ GeV (left-most orange ``Tevatron'' curve) the chargino becomes lighter than the stop, and the discussed channel is no longer effective.

Besides searches for a light stop, the LHC will abundantly produce not only gluinos, but charginos and neutralinos as well. The corresponding LHC phenomenology was recently addressed in a similar setting, namely, the focus point region of minimal supergravity~\cite{Baer:2005ky}. Here, however, the gluino cascade decay channels -- in the presence of a light stop  --  would be significantly different, even though, as in the focus point region, all other scalars are very heavy, since the gluino would largely decay into the lightest stop rather than in neutralinos/charginos. An alternative search channel analyzed in Ref.~\cite{Baer:2005ky} is the clean trilepton signature originating from the decays of the pair $\widetilde \chi^\pm_1\ \widetilde \chi^0_2$ following the reactions $\widetilde \chi^\pm_1\rightarrow l\overline\nu \chi$ and $\widetilde \chi^0_2\rightarrow l\bar l\chi$. In this channel the LHC reach will extend up to $M_1\lesssim220$ GeV, where a sufficiently large flux of  supersymmetric particles  can be produced, provided the mass splitting between the two decaying particles and the LSP is large enough (roughly, $m_{\widetilde \chi^\pm_1}-m_\chi\gtrsim 40$ GeV). In Fig.~\ref{fig:colliders} we indicate the contours where the two above mentioned conditions (light enough charginos/neutralinos and large enough mass splitting) are both matched.

Finally, the reach of a future international linear collider (ILC) in the light-stop scenario has been recently addressed in Ref.~\cite{Carena:2005gc}. Even with a moderate integrated luminosity, say 10 ${\rm fb}^{-1}$, the discovery reach of a $\sqrt{s}=500$ GeV ILC for production of light stops decaying into $c\chi$ extends up to very small neutralino-stop mass splittings, and a very precise determination of superpartner masses and mixing angles has been shown to be feasible \cite{Carena:2005gc}. This channel, however, implies a lighter chargino heavier than the lightest stop, and hence is possible in the left part of Fig.~\ref{fig:colliders} (roughly out to the left-most ``Tevatron'' curve). In the higgsino-like LSP region (bottom right), the stop would decay into $b\widetilde\chi^+_1$, a channel that has not yet been studied for the ILC but that could also give distinct signatures. For definiteness, we show in Fig.~\ref{fig:colliders} the contours at $m_{\widetilde t_1}-m_\chi=10$ GeV. Further, charginos and heavier neutralinos would be abundantly produced in that region, and one can expect that the ILC energy will be close to the lightest chargino mass threshold ($m_{\widetilde\chi^\pm_1}<\sqrt{s}$). Since the mass splitting between the chargino and the LSP can be very small in the pure higgsino LSP region, we show a conservative prospective reach contour for a 500 GeV ILC in Fig.~\ref{fig:colliders}, extending it to chargino masses as heavy as only 225 GeV.

In the case of an anomaly mediation gaugino spectrum, stops will always decay into charginos and $b$ quarks, a channel lying beyond the Tevatron reach. We expect a collider reach for the LHC very similar to that discussed in the framework of the minimal anomaly mediated supersymmetry breaking scenarios in Ref.~\cite{Baer:2000bs}. For the ILC, on the other hand, we estimate a reach relatively close to the kinematic thresholds in the chargino and stop pair production channels.

Our summary of the collider capabilities in regions of the MSSM compatible with resonant EWB shows a high degree of complementarity between hadron colliders and an ILC. We illustrate this showing the BAU-compatible regions with a light-blue shading, for $|\sin\phi_\mu|=0.5$ (smaller phases would produce contours contained inside those we show (see Fig.~\ref{fig:yb}), while larger phases would be, in general, in conflict with the electron EDM (see Fig.~\ref{fig:edm})). The bino-like region of the parameter space can be probed at the Tevatron, in a very narrow corner of parameter space; the LHC might detect light stops, again in the bino-like region only, in the same-sign top channel, and the trilepton signature will cover almost the entire region where the $\mu\sim M_2$ resonance lies. If instead nature lives in the lower $M_1\sim\mu$ resonance corresponding to neutralino-driven EWB, then the discovery of superpartners at the LHC could be extremely challenging, but even a 500 GeV ILC would abundantly produce charginos and stops, and allow for detailed studies of the superpartners' properties.

\section{Conclusions}\label{sec:conclusions}

In this work, we have analyzed the implications of existing and future EDM and DM searches for the viability of EWB in the MSSM. We have concentrated on scenarios of resonant -- or nearly resonant -- EWB, for which the requirements on the CP violating phases are least demanding and the EDM results least constraining. These scenarios correspond to the supersymmetric higgsino mass parameter $\mu$ having magnitude close to that of either of the soft, electroweak gaugino masses, $M_{1,2}$. The presence of these resonant effects leads to a characteristic \lq\lq double funnel" contour in the $\mu$-$M_i$ planes corresponding to the regions consistent with the observed baryon asymmetry. In particular, we note that for $\mu\sim M_1$, resonant EWB is driven entirely by neutralinos, and that this scenario is presently not ruled out by other experiments. In models of SUSY-breaking mediation that incorporate GUT-scale gaugino mass unification, one of these neutralinos can also be the LSP. On the other hand, for $\mu\sim M_2$, both charginos and neutralinos contribute to $Y_B$, and only in AMSB-type models with $M_2 < M_1$ can one of these neutralinos also be the LSP.

For either case, the mass parameters $\mu$ and $M_i$ should lie in the sub-TeV domain in order to produce a sufficiently large BAU, and -- modulo remaining uncertainties in $Y_B$ calculations -- the minimal value of $|\sin\phi_\mu|$ for successful EWB is  $\sim 10^{-2}$. Additional constraints are imposed by the results of EDM and DM searches. In the case of EDMs, results for the electron lead to the most stringent constraints\footnote{A feature that does not hold in general for other models of new CP-violation.}. Consequently, we compared the one- and two-loop $d_e$ constraints on the MSSM parameter space compatible with resonant EWB. We showed that one- and two-loop effects are comparable when the slepton masses lie in the range 3-10 TeV, and that consistency with EWB requires the slepton masses to be larger than a TeV. Moreover, in portions of the MSSM parameter space consistent with the observed BAU, two-loop effects constrain the CP violating phase $\phi_\mu$ to be smaller than its maximal value; we expect these contributions to be larger than $10^{-28}$ e-cm, and therefore accessible to future experimental scrutiny.

When analyzing the implications for neutralino DM, we find that in the regions wherein the LSP contributes to resonant EWB,  the DM particle is a largely mixed gaugino-higgsino state. Assuming $M_1\ll M_2$, we showed that there exist portions of the parameter space in which the observed $Y_B$ can be resonantly produced and the observed LSP relic abundance can arise from thermal production. However, in order to have neutralino-driven EWB, additional, non-thermal production or cosmological enhancement mechanisms must be invoked. In contrast, for  $M_2\ll M_1$, then thermal neutralinos  are always under-produced by factors ranging from $\sim 30$ to $\sim 300$, so that they must always undergo a cosmological enhancement or must be supplemented with a non-thermally produced population.

In regions of the MSSM parameter space for which the LSP participates in resonant EWB, we expect production of large fluxes of energetic solar neutrinos induced by the capture and annihilation of neutralinos in the Sun.  The absence of evidence for such neutrinos in data collected by the SuperKamiokande collaboration leads to strong constraints on these regions of the MSSM parameter space. Future ${\rm km}^3$ neutrino telescopes will likely be sensitive to the flux of neutrinos from the Sun expected from most of EWB-compatible MSSM models where the source of CP violation resides in the gaugino/higgsino sector, even relaxing the requirement of a light stop and assuming that other mechanisms -- such as extended Higgs sectors -- induce a sufficiently strong first-order EW phase transition. We also showed that the EWB scenario will be largely testable at future ton-sized direct detection experiments, and studied the dependence of both neutrino telescope rates and direct detection rates on the CP violating phase.

Finally, future collider experiments will provide additional probes of the parameter space relevant to EWB, with a considerable degree of complementarity between the LHC and ILC reaches. In particular, the LHC is expected to cover most of the EWB parameter space if the LSP does not contribute to resonant EWB, whereas the ILC will be able to probe EWB-DM connected scenarios as well. 


\vspace*{1cm}
\noindent{{\bf Acknowledgments} } \\
\noindent This work was supported in part U.S. Department of Energy contracts FG02-05ER41361 and DE-FG03-92-ER40701,  N.S.F. award PHY-0071856, and NASA contract NNG05GF69G. VC was also supported by the Caltech Sherman Fairchild fund.


\end{document}